\documentclass[12pt]{article}
\usepackage{graphicx}
\usepackage{lscape}
\usepackage{eurosym}
\bibstyle{plain}

\newcommand{\be}{\begin{equation}}
\newcommand{\en}{\end{equation}}

\newcommand{\PP}{{\mathord{I\kern -.33em P}}}
\newcommand{\EE}{{\mathord{I\kern -.33em E}}}
\newcommand{\RR}{{\mathord{I\kern -.33em R}}}

\newcommand{\ea}{\end{eqnarray}}
\newcommand{\ba}{\begin{eqnarray}}
\newcommand{\ean}{\end{eqnarray*}}
\newcommand{\ban}{\begin{eqnarray*}}

\begin{document}

\title{X-Value adjustments: accounting versus economic management perspectives} 

\author{Alberto Elices\thanks{Head of XVA Model Validation, Bank Santander, Av. Cantabria s/n, 28660 Boadilla del Monte, Spain, {\em aelices@gruposantander.com}.} } 
\date{\today}
\maketitle

\begin{abstract}
{\small 
This paper provides a mathematical framework based on the principle of invariance \cite{2016_InvariancePrinciple} to classify institutions in two paradigms according to the way in which credit, debit and funding adjustments are calculated: accounting and management perspectives. This conceptual classification helps to answer questions such as: In which paradigm each institution sits (point of situation)? Where is the market consensus and regulation pointing to (target point)? What are the implications, pros and cons of switching perspective to align with future consensus (design of a transition)? An improved solution of the principle of invariance equations is presented to calculate these metrics avoiding approximations and irrespective of the discounting curve used in Front Office systems. The perspective is changed by appropriate selection of inputs always using the same calculation engine. A description of balance sheet financing is presented along with the justification of the funding curves used for both perspectives.

\textbf{Disclaimer}: the views expressed in this article are exclusively from the author and do not necessarily represent the views of neither Bank Santander nor its affiliates.
}
\end{abstract}

\section{Introduction}
\label{sec:Introduction}

Counterparty Credit Risk in financial derivatives and funding cost have increasingly become topics of research since the credit crisis in 2008. After the default of Lehman Brothers the assumption that financial institutions could not default was no longer accepted. This hypothesis has been introduced in the existing pricing framework by using the previous risk free pricing plus a number of adjustments: the CVA (Credit Value Adjustment) to take into account the default of the counterparty and the DVA (Debit Value Adjustment) to account for the default of the institution from which the pricing is carried out (from now on ``the bank''). To reduce counterparty credit risk, collateralization was generalized among big institutions in the following years.

Collateralization came for the over-the-counter market in the form of variation margin to cover the liable mark-to-market in case of default. Additionally, regulators have fostered closing operations through central clearing counterparties which require posting an additional initial margin to cover the market variation during the margin period of risk (usually two weeks) from counterparty default to liquidation. Regulators have also required higher amounts of initial margin posting for over-the-counter markets to foster closing operations with central clearing counterparties. Collateral can be cash or other type of liquid assets (e.g. bonds, Equity, etc). This mechanics required funding this collateral and the need for additional adjustments such as the FVA or Funding Value Adjustment, to account for the funding costs and benefits of collateral or the MVA, Margin Value Adjustment, to address the cost of funding of initial margin. Finally, the last adjustment which has been proposed is the Capital Value Adjustment or KVA which accounts for the funding costs of capital required by regulators. All these adjustments are all denominated under the acronym XVA.

Since 2008, a continuous debate has taken place among institutions and regulators on how these adjustments should be calculated and reported to comply with accounting standards and to properly foster internal management. After more than a decade, the CVA has settled down as an accepted adjustment, DVA and FVA are still under debate and the rest are still far from being standardized.

Two major paradigms have been discussed around XVA and in particular about FVA since 2012. The first one, mainly advocated by Hull and White in \cite{2012_FVA_debate}, considers that the value of a firm defined as the shareholder and creditor value does not depend on the funding strategy of the institution according to Modigliani-Miller theorem \cite{1958_MMT}. Therefore the FVA should be zero, derivative prices are symmetric (as so CVA and DVA are) and obey the law of one price. This paradigm can well be used for accounting fair value. Burgard and Kjaer study the impact of derivatives and their corresponding funding positions in the balance sheet upon the issuer default (see \cite{Burgard2012_Balance} and \cite{Burgard2017_Symm_rates}). They have also proposed ways to mitigate this impact so that funding adjustments can be dropped.

The second paradigm advocates for the inclusion of the FVA in derivative prices which are charged to final customers in order not to reduce shareholder value for the cost of collateral funding of hedges associated with unsecured operations. This paradigm implies wealth transfer from shareholders to senior creditors as the valuation of the extra funds required from the latter to fund derivatives (monetized when the bank defaults) equals the funding value adjustment charged to customers (see \cite{2014_Acc_FVA_Albanese_Andersen}). This paradigm is more aligned with what practitioners and institutions consider how the management should be carried out. On the other hand, the funding benefits provided by FVA overlap with DVA profit on bank liabilities when funding benefits are valued at the internal transfer funding rate of the bank which accounts for the bank credit spread, already included in DVA (see for instance \cite{Gunnenson2014}). In order to avoid this double counting, some practitioners prefer to drop DVA and consider only FVA and others to preserve both but only the liquidity component of the internal funding spread in the FVA. In fact, this latter approach is a point in between both paradigms because avoiding double counting between DVA and FVA allows both adjustments to live together providing a symmetric price between the counterparty and the bank. This price would be compatible with the law of one price and the accounting regulation (the essence of the first paradigm).

This paper does not provide additional support for either paradigm but simply pursues to accommodate both approaches into a common framework which can be ``configured'' to align with the first paradigm, the ``accounting perspective'' or with the second paradigm, the ``management perspective''. The choice of perspective will depend on the decision of the senior management of each institution which will progressively accommodate with the evolution of the market consensus (see section \ref{sec:ManagingPerspective}) and the regulatory requirements (see section \ref{sec:AccountingPerspective}).

This paper has four major contributions. The first one provides a framework, based on the principle of invariance \cite{2016_InvariancePrinciple}, which conceptually allows classifying institutions into these two paradigms in relation with credit, debit and funding value adjustment calculation: an accounting and an economic management perspective with intermediate transitions among them (see sections \ref{sec:AccountingPerspective} and \ref{sec:ManagingPerspective}). This classification allows a conceptual identification of which perspective each institution sits on, where is the market consensus settling down (section \ref{sec:ManagingPerspective} summarizes the consensus surveys \cite{2017_SurveyFVAasymmetry}, \cite{2018_SurveyXVA} and \cite{Nov2019XVASurveyPwC}), to which perspective the regulation is pointing to (section \ref{sec:ComparisonRegulation}), what are the implications, advantages and disadvantages of moving from one perspective to the other (see section \ref{sec:ComparisonPerspectives}) and the prospective difficulties of institutions to adjust to the tendency of market consensus and regulation to the management perspective (see section \ref{sec:Transition}). This conceptual mindset provides a route guide to realize about the current point of situation, the target point, how to design a transition from one to the other and allowing reporting the stage of completion of this transition.

The second contribution provides an improved solution of the CVA/DVA equation with funding based on the mathematical framework of the principle of invariance \cite{2016_InvariancePrinciple} (it is reviewed in sections \ref{sec:InvariancePrinciple} and \ref{sec:InvariancePrincipleDefault}). This improved solution eliminates a circular dependence which appears in the original equation (see sections \ref{sec:NotCircularDep} and \ref{sec:NotCircularDepDefault}). This avoids some approximations which indeed violate the condition of the principle of invariance given by equation (\ref{eq:funding_eq}). In addition, this solution calculates the XVA adjustments in terms of the exposures discounted with whatever rate is available in the system as presented in equations (\ref{eq:fundcr_Vmx}) to (\ref{eq:CDVA_Vmx}) where no conversion or approximation needs to be carried out.

The third contribution is that the mathematical framework is common for both perspectives. Each of them is defined by a choice of a set objects of selection (see section \ref{sec:FrameworkObjectsSelection}). This allows a smooth transition from one perspective to another without having to implement changes in the calculation engine.

Finally, the fourth contribution justifies how funding rates can be estimated under the assumption that the Financial area is not a profit center. For the accounting perspective the internal funding rate is related with the average liquidity spread of the bonds issued by the institution. On the other hand, for the management perspective the internal funding rate is an average yield of the issued bonds.

The paper starts with a tour around the management of balance sheet financing from a descriptive point of view. This includes a description of the functions of the financial area and the financial management control (section \ref{sec:FinancialArea}), the bond issuance department (section \ref{sec:IssuanceDep}), the short term desk (section \ref{sec:ShortTermDesk}) and the Securities Financing desk (section \ref{sec:SecuritiesFinancing}). Section \ref{sec:ExampleFTPrate} illustrates with a simple example how to estimate the internal FTP (Fund Transfer Pricing) rate based on the bond issuance activity and various hypotheses which will be thereafter aligned for the derivation of the appropriate FTP rate depending on the perspective adopted (either accounting or management).

Section \ref{sec:InteractionCVADVAFVA} discusses the interaction among credit, debit and funding value adjustments in terms of collateral and concludes that the best way to manage collateral is to reach a balance between the positions which generate CVA (unsecured assets in favour of the bank) and the positions which generate DVA (unsecured liabilities in favour of the counterparty). This section is a bridge from the first descriptive part of the paper and the second part with the presentation of the mathematical framework, the two perspectives and their interpretation and implications.

The presentation of the mathematical framework in section \ref{sec:FrameworkObjectsSelection} is introduced by a review and an interpretation of the principle of invariance with and without default in sections \ref{sec:InvariancePrinciple} and \ref{sec:InvariancePrincipleDefault}. Sections \ref{sec:NotCircularDep} and \ref{sec:NotCircularDepDefault} present the improved solution of the invariance equations\footnote{Developed by Jérôme Maetz, Head of XVA Front Office quant team, Santander Bank.\label{fnote:Jerome}} of section \ref{sec:InvariancePrincipleDefault} with and without default to calculate credit, debit and funding value metrics avoiding approximations and irrespective of the discounting curve used in Front Office systems.

Sections \ref{sec:AccountingPerspective} and \ref{sec:ManagingPerspective} justify and motivate the accounting perspective (regulation) and management perspective (hedging and market consensus). Each one is defined by making choices for the objects of selection introduced in the mathematical framework (section \ref{sec:FrameworkObjectsSelection}). Both perspectives are compared in section \ref{sec:ComparisonPerspectives}.

Section \ref{sec:ComparisonRegulation} analyzes the FRTB-CVA regulation and justifies why it is pointing to the management perspective. Finally, section \ref{sec:Transition} analyzes the implications of the transition from the accounting to the management perspective and explains why there are institutions which may be reluctant to change until the market consensus is clearly settled. Section \ref{sec:Conclusions} ends up with some conclusions.

The mathematical derivations are not included in the body of the paper but left for the appendices. Appendix \ref{sec:DiscEqDerivation} shows the derivation of the relation between two prices discounted with two different curves. Appendix \ref{sec:Invariance_default} shows the derivation of the equations of the principle of invariance with default risk where the invariance rate which can be arbitrarily chosen has not yet been selected. Finally, appendix \ref{sec:XVAimproved} shows the derivation of the improved solution which breaks the circular dependence and expresses XVA metrics in terms of the discounting used in the internal Front Office systems rather than the invariance rate.

\section{Management of balance sheet financing}
\label{sec:BalanceSheet}

In order to properly understand how the balance sheet management of a bank is carried out, this section briefly reviews the main areas and departments in which this function is located: the financial area which provides structural financing to the bank, the Issuance Department which allows dynamically fitting the financial requirements of the bank, the Short Term desk to manage the daily closeout of cash flows and the Securities Financing desk which optimally manages collateral. Section \ref{sec:ExampleFTPrate} provides a worked example which provides an approxiamtion of how the internal Funding Transfer Pricing may be calculated given some hypotheses.

\subsection{Financial area}
\label{sec:FinancialArea}

The main purpose of the Financial area of a bank is to provide structural short and long term financing to the bank. The structural financing is needed to support the ordinary businesses of the bank and other corporate operations for which liquidity is needed.
\begin{itemize}
	\item \textbf{Ordinary businesses of the bank}: examples of this category are credit lending (e.g. provide credit to retail/institutional customers, mortgages, etc) derivatives, which also involve a part of liquidity lending or borrowing.
	\item \textbf{Corporate operations}: examples of corporate operations can be to acquire or increase participatio in other entities, real state operations or mergers.
\end{itemize}

The \textbf{building blocks} of the Financial area are the various areas and departments which provide the services to allow raising the structural liquidity that a bank needs in each moment:
\begin{itemize}
	\item \textbf{Financial management control}: this department calculates and distributes the internal financing rates or FTP (Fund Transfer Pricing) rate. As part of the strategic vision or the bank, some businesses may be fostered providing lower financing rates and others mitigated with higher financing rates. This is of course a decision of the senior management of the bank and he actual implementation of this strategic views are carried out by this department.
	\item \textbf{Issuance department}: this department issues corporate paper to provide additional liquidity to close the balance of the bank so that all financing needs are equal to the financial sources of the bank. In general, the bank is usually organized to lack liquidity. Therefore, the Issuance Departement keeps constantly issuing corporate paper.
	\item \textbf{Short Term desk}: provides liquidity to daily close-out cash payments of every desk in every currency. The surplus or extra liquidity needed is first given or provided by the Interbank lending/borrowing and thereafter by the internal liquidity of the bank for which regulatory ratios must be satisfied.
	\item \textbf{Securities Financing desk}: this desk allows short term asset and collateral optimization to maximize return and collateral reuse when collateral is rehypothecable. 
\end{itemize}

The management of the Financial area is carried out by collecting all the assets and liquidity from every business of the bank. The main ways the bank can raise money are the deposits or cash coming from the customers, Equity and debt issuance or Central banks. All sources of liquidity are deposited in a common account joining liquidity provided and consumed by every business.
\begin{itemize}
	\item \textbf{Equity issuance}: Equity issuance is carried out only from time to time to raise internal capital from shareholders. Preparing an Equity issuance is a relevant event for which many market participants needs to be coordinated (that is why Equity issuance is occasional). There are examples in which there is a continuous Equity issuance process such as the payment of dividends in form of stock or stock dividends. The bank does not pay dividends and in exchange issues shares to pay its shareholders.
	\item \textbf{Businesses of the bank}: most of the financing a bank obtains comes from customer deposits from their retail business. The retail business is carried out through the bank agencies and more lately through the digital customer relationship. In addition, commercial banking (institutional customers) or corporate banking (corporates) may be other ways of raising financing (sometimes they request for liquidity) for the bank through special bank analysts who are specialized in bigger customers (e.g. private banking).
	\item \textbf{Central banks}: In some situations Central Banks or Central Bank agencies may offer packages of cheaper financing than Capital markets. Examples are Targeted Longer-Term Refinancing Operations (TLTRO) from the European Central Bank or excess of liquidity in various currencies lent by agencies such as SAMA (``South Arabian Monetary Authority'').
	\item \textbf{Debt issuance}: there is a big range of debt which may be issued by banks. Bonds are issued for medium and long terms with a wide range of seniority depending on the type of credit quality. Commercial paper is usually issued for short term and covered bonds are bonds backed by mortgages. The issuance process is carried out continuously to match lack of liquidity and renew expired debt.
\end{itemize}

The budget for the finance area is built one year ahead collecting the needs and sources of liquidity from every area. This budget is calculated from a conceptual basis. However, there are also financing operations closed internally from the Financial area for a given time horizon with various businesses of the bank as long as they happen. This operations officially transmit liquidity surplus or requirements so that the Financial area may coordinate with the Issuance department the provisioning of these extra funds. There are also situations in which the Financial area requests the desks (e.g. the Equity desk) to coordinate a campaign to raise liquidity if they anticipate that the Issuance department may not cope with raising the whole amount of liquidity needed.

\subsection{Issuance department}
\label{sec:IssuanceDep}

The Issuance department is the heart of the dynamic liquidity management. This department closes out sources and needs of medium and long term liquidity. The main objectives of the Issuance department are maintaining regulatory ratios within appropriate levels, providing adjustments to the liquidity budget (based on internal term operations closed with the Financial area) and re-finance expiring debt to modulate in terms of current liquidity needs or strategic views, the total amount of debt outstanding.

The regulatory ratios which are under the scrutiny of the Issuance department so that they are kept within appropriate levels (usually higher than 100\%) are the following:

\begin{itemize}
		\item \textbf{LCR} (Liquidity Coverage Ratio): liquid assets over payments in the coming 30 days. This is the most basic short term ratio. When Lehman Brothers collapsed in September 2008, part of the issue was that they were getting financing from the Interbank market and it got completely dried (no more borrowing or lending). This ratio was introduced to avoid such situations.
		\item \textbf{NSFR} (Net Stable Funding Ratio): ratio of stable funding (e.g. liabilities beyond 1 year maturity). This ratio fosters liabilities which do not expire in such a short term as the LCR.
		\item \textbf{TLAC} (Total Loss Absorbing Capacity): this ratio assures enough Equity and bail-in debt to pass losses to investors to avoid a government bailout. This ratio is to avoid governments to have to take money from tax payers to rescue financial institutions.
		\item \textbf{MREL} (Minimum Requirement for own funds and Eligible Liabilities): this ratio is more related with the degree of resolvability to absorb losses and restore capital position during and after a crisis. This ratio is a post default orderly management measure.
\end{itemize}

In order to maintain these regulatory ratios beyond 100\%, the Issuance department may have to issue various types of debt with varying maturity and credit seniority. The hierarchy of asset quality and the corresponding capital impacts are the following:
\begin{itemize}
	\item \textbf{Equity shares, derivative profit-and-loss not including DVA}: these are part of the CET1 (Common Equity Tier 1). Equity shares and traded derivatives are the most liquid type of assets which can be liquidated in the market at once. The reason why DVA and the rest of own credit adjustments cannot be considered part of the Common Equity Tier 1 capital arises from \cite{DeductionDVAfromCET1} in which the Basel Committee on Banking Supervision resolves not to recognize this type of asset as Common Equity for the profit it provides can only be realized in case of default of the institution.
	\item \textbf{Commercial paper}: this is a liquid debt issued by the bank to obtain short term liquidity. In many situations, commercial paper is allowed to be recognized as part of CET1 capital. However, there are situations in which it is not.
	\item \textbf{Preferred participations}: this type of debt is  part of what is called the AT1 (Additional Tier 1 capital). This is the following layer of CET1. It is not as liquid as CET1 but it is still not considered tier 2 capital.
	\item \textbf{Subordinated debt}: this type of debt has the lowest priority within the hierarchy of creditors in case of liquidation or bankruptcy. Subordinated debt holders get paid just before stock holders. This type of debt is part of T2 (Tier 2) capital, the second layer according to Basel regulation. It is also counted as part of TLAC ratio.
	\item \textbf{Senior non-preferred debt}: this type of debt has higher priority in the creditor hierarchy than subordinated debt. It is counted for the calculation of the TLAC ratio.
	\item \textbf{Senior preferred or unsecured debt}: this is the debt at the top of the hierarchy of creditor priority.
	\item \textbf{Covered Bonds}: these bonds are usually over collateralized by a pool of mortgages and represent the closest debt to risk-free.  
\end{itemize}

As a conclusion, the Issuance department allows raising funds to dynamically manage deviations in the financing budget of the bank. Budget deviations are formally acknowledged through internal term operations between various areas of the bank and the Financial area. The wide range of debt issued is optimized to maintain in appropriate levels a number of regulatory ratios. The debt issuance is a continuous process which renews expired debt and can be modulated in the long term to shape the structure of the total debt outstanding. Depending on the liquidity of the market, the Issuance department may have to change the strategy ahead of time to be able to cope with the liquidity constraints of the market.

\subsection{Short term desk}
\label{sec:ShortTermDesk}

The short term desk daily closes out all the cash positions in every currency for the legal entity. The way in which the short term desk operates is summarized in the following points:

\begin{itemize}
	\item When the bank needs to carry out payments and it does not have money, the short term desk borrows it on a daily basis from the Interbank market to accomplish all payments. When there is surplus of cash, they lend it to the Interbank market. Daily overnight interbank lending/borrowing is carried out at OIS rate to match excess/lack of liquidity.
	\item When the liquidity of the Interbank market dries out, the short term desk takes the liquidity from the common account where all sources of financing are collected. Daily payments can never be failed as this would technically trigger a default. Therefore, the common account must be there as a backup to never run out of liquidity.
	\item Shortage in the common account triggers debt issuance through regulatory ratios. When the short term desk starts to take liquidity from the common account, this deteriorates the regulatory capital ratios. Therefore, the Issuance department will need to issue more debt to restore healthy levels for these ratios.
	\item Liquidity in currencies other than the mayor currency of the bank (e.g. EUR for European, USD for North American or JPY for Japanese banks) is obtained by borrowing in the major currency at the interbank market OIS rate and exchanging it to the required currency by means of foreign exchange spot or forward (very short term) operations. Therefore, the effective rate is the OIS rate plus the cross currency basis spread. Penetration in other currencies could make it cheaper to obtain liquidity in currencies different from the major one. However, in practice, most of the liquidity in foreign currencies is obtained through spot/forward operations.
\end{itemize}

\subsection{Securities Financing desk}
\label{sec:SecuritiesFinancing}

This desk optimizes a pool of assets and collateral to maximize return and match LCR (Liquidity Coverage Ratio) and collateral requirements at the lowest possible cost. The tasks and machinery carried out by this desk are the following:
\begin{itemize}
	\item \textbf{Asset rotation}: this process consists of leveraging by exchanging high for low quality assets for a pick-up spread and de-leveraging by getting back the high quality assets by paying the spread back. The leveraging process allows reusing assets received as collateral when they are not needed to get some extra return and the de-leveraging process allows to rescue some assets which at a given point in time may be needed to post as collateral or to increase regulatory ratios. Leveraging increases return but reduces liquidity (the liquid assets are temporarily sold or exchanged) and de-leveraging improves liquidity and LCR ratio but reduces return.
	\item \textbf{Asset lending/borrowing}: agreed among institutions for received or paid pick-up spread from funds or banks. This is part of the machinery developed by the Securities Financing desk to increase support and coverage from other institutions in case collateral or other assets may be needed in the present or future.
	\item \textbf{Internal operations}: closed with desks or other areas of the bank to get or request liquidity for specific terms. Sometimes the derivative desks are fostered to get liquidity when the market situation may not be favorable to obtain it by promoting derivative campaigns to customers through the retail network.
	\item \textbf{Pool of assets}: the Securities Financing desk has access to an extensive pool of assets such as bonds, letters, commercial paper, Equity assets, etc. This pool of assets has to be optimized so that the bank can post appropriate collateral or have assets available for regulatory ratio computation. This desk is more dynamic than the Issuance department and may help it to maintain some of the regulatory ratios within appropriate levels on a periodic basis.
	\item \textbf{Collateral reuse}: the asset rotation by the lending/borrowing machinery allows raising the ratio of collateral reuse up to more than 80/90\%. This is for the collateral which is re-hypothecable and therefore can be reused. Segregated or non-rehypothecable collateral cannot be touched and it is usually kept inside a custodian entity. 
	\item \textbf{Typical operations}: the range of operations which the Securities Financing desk usually closes includes repo and reverse repo transactions, total return swaps, security lending, futures and some options.
\end{itemize}

Some institutions reuse extra collateral coming out of liability porfolios by posting it as initial margin. This is an example of how to reuse collateral in an efficient way.

\section{Estimation of Fund Transfer Pricing rate}
\label{sec:ExampleFTPrate}

This section shows a simple and intuitive approximation to determine the financing costs of a bank based on the bond issuance process under two different hypotheses. The first one assumes that the bank can default and only the recovery of the outstanding debt will be paid back to bond holders. The second one will consider that the bank cannot default.

Consider a bank $B$ which issues bonds on dates $t_i$ expiring on $T_i$ with notional $F_{i,t}  = F_i  \cdot {\bf 1}_{\left\{ {t_i  < t < T_i } \right\}}$ and valued on date $t$ at a price $B_{i,t}^B$:

\begin{equation}
B_{i,t}^B  = F_{i,t} \left\{ {\int_t^{T_i } {p_{t,s}^r Q_{t,s}^B r_{t_i ,s}^B ds}  + p_{t,T_i }^r Q_{t,s}^B  + \int_t^{T_i } {R_s^B p_{t,s}^r Q_{t,s}^B \lambda _{t,s}^B ds} } \right\}
  \label{eq:Bit_b}
\end{equation}

where $r_{t_i,t}^B$ is the bond continuous coupon, $R_s^B$ and $\lambda_t^B$ are the recovery rate and term structure of the instantaneous hazard rate on $t$ of bank $B$. Equation (\ref{eq:Pts_Qts}) presents the values of $p_{t,s}^r$ and $Q_{t,s}^B$ which are the discount factor (with instantaneous risk free forward rate $r_{t,s}$) and survival probability from $t$ to $s$.

\begin{equation}
\begin{array}{*{20}c}
   {p_{t,s}^r  = \exp \left( { - \int_t^s {r_{t,u} du} } \right)} & {} & {Q_{t,s}^B  = \exp \left( { - \int_t^s {\lambda _{t,u} du} } \right)}  \\
\end{array}
  \label{eq:Pts_Qts}
\end{equation}

The liquidity obtained from issued bonds is now transferred to the derivative desks through an internal transfer bond issued by the Financial area and sold to the desk. This bond is issued at a price $B_{i,t}$ with full recovery. The reason to consider full recovery is because both the desk and the Financial area belong to the same bank and in case of default, all the assets of the bank will be put together to face the recovery payments. Therefore, it does not make sense that the desk pays only a recovery to the Financial area. Instead, the assets held by both are used to pay the recovery to the bond holders. The price of the internal bond is given by equation (\ref{eq:Bit}):

\begin{equation}
B_{i,t}  = F_{i,t} \left\{ {\int_t^{T_i } {p_{t,s}^r Q_{t,s}^B r_{t_i ,s}^{FTP} ds}  + p_{t,T_i }^r Q_{t,s}^B  + 1 \cdot \int_t^{T_i } {p_{t,s}^r Q_{t,s}^B \lambda _{t,s}^B ds} } \right\}
  \label{eq:Bit}
\end{equation}

The Financial area balance on date $t$ is given by equation (\ref{eq:FAt}) assuming positive bond price ($B>0$) for assets and negative bond price ($B<0$) for liabilities of the Financial area. With this criterion, the internal bonds are positive (the Financial area has to get paid from the desks) and issued bonds are negative (the Financial area has liabilities with the bond holders). The term $T= \max \left\{ {T_i ,{\rm  }i = 1 \cdots NB } \right\}$ refers to the longest expiry of issed debt and NB refers to the Number of Bonds issued by the bank.

\begin{equation}
\begin{array}{l}
 FA_t  = \sum\limits_{i = 1}^{NB} {B_{i,t}  - } \sum\limits_{i = 1}^{NB} {B_{i,t}^B }  =  \\ 
 = \int_t^{T} {p_{t,s}^r Q_{t,s}^B } \left\{ {\sum\limits_{i = 1}^{NB} {F_{i,t} \left( {r_{t_i ,s}^{FTP}  - r_{t_i ,s}^B  + \left( {1 - R_s^B } \right)\lambda _{t,s}^B } \right)} } \right\}ds \\ 
 \end{array}
  \label{eq:FAt}
\end{equation}

Assuming that the total debt issued is $F_t  = \sum\limits_{i = 1}^{NB} {F_{i,t} }$, the weight of each bond issuance is $w_{i,t}  = F_{i,t} /F_t$ and the CDS spread of the bank on date $t$ is $CDS_{t,s}^B  = \left( {1 - R_s^B } \right)\lambda _{t,s}^B$, the balance given by equation (\ref{eq:FAt}) turns into equation (\ref{eq:FAt_CDS}). For CDS quotation purposes, sometimes the recovery rate is set to a given standard value (e.g. $R = 0.4$). This would mean that $\lambda^B_{t,s}$ has the information of both the recovery ratio and the default intensity. However, for the purpose of this discussion, $R_s^B$ is the varying recovery rate at time $s$. See that equation (\ref{eq:FAt_CDS}) adds and subtracts the CDS spread, $CDS_{t_i}^B$, at the issuance time $t_i$:

\begin{equation}
FA_t  = F_t  \cdot \int_t^T {p_{t,s}^r Q_{t,s}^B \sum\limits_{i = 1}^{NB} {w_{i,t} \left( \begin{array}{l}
 r_{t_i ,s}^{FTP}  - r_{t_i ,s}^B  + CDS_{t,s}^B  \\ 
  + CDS_{t_i,s}^B  - CDS_{t_i,s}^B  \\ 
 \end{array} \right)ds} } 
  \label{eq:FAt_CDS}
\end{equation}

Now the yield of the issued bond can be decomposed into OIS rate, $r_{t_i ,s}^{OIS}$, plus CDS spread and liquidity spread, $l_{t_i ,s}^B$, according to equation (\ref{eq:rti_b}):

\begin{equation}
r_{t_i ,s}^B  = r_{t_i ,s}^{OIS}  + \left( {1 - R_{t_i }^B } \right)\lambda _{t_i ,s}^B  + l_{t_i ,s}^B  = r_{t_i ,s}^{OIS}  + CDS_{t_i ,s}^B  + l_{t_i ,s}^B 
  \label{eq:rti_b}
\end{equation}

If equation (\ref{eq:rti_b}) is replaced in equation (\ref{eq:FAt_CDS}), the expression of equation (\ref{eq:FA_t_CDS}) is obtained where the $CDS_{t_i}^B$ inside $r_{t_i,s}^B$ cancels with the one from equation (\ref{eq:FAt_CDS}).

\begin{equation}
FA_t  = F_t  \cdot \int_t^T {p_{t,s}^r Q_{t,s}^B \left\{ {\sum\limits_{i = 1}^{NB} {w_{i,t} \left( \begin{array}{l}
 r_{t_i ,s}^{FTP}  - r_{t_i ,s}^{OIS}  - l_{t_i ,s}^B  \\ 
  + CDS_t^B  - CDS_{t_i }^B  \\ 
 \end{array} \right)} } \right\}ds} 
  \label{eq:FA_t_CDS}
\end{equation}

Finally, for the Financial area not to be a profit center, namely $FA_t = 0$, the following average FTP rate, $\bar r_t^{FTP}$, follows, where the rest of average values are provided by equation (\ref{eq:rav}):

\begin{equation}
\bar r_{t,s}^{FTP}  = \bar r_{t,s}^{OIS}  + \bar l_{t,s}^B  + \overline {CDS} _t^B  - CDS_t^B \approx \bar r_{t,s}^{OIS}  + \bar l_{t,s}^B
  \label{eq:rtF}
\end{equation}

\begin{equation}
\begin{array}{*{20}c}
   {\bar r_{t,s}^{FTP}  = \sum\limits_{i = 1}^{NB} {w_{i,t} r_{t_i ,s}^{FTP} } } & {\bar r_{t,s}^{OIS}  = \sum\limits_{i = 1}^{NB} {w_{i,t} r_{t_i ,s}^{OIS} } }  \\
   {\bar l_{t,s}^B  = \sum\limits_{i = 1}^{NB} {w_{i,t} l_{t_i ,s}^B } } & {\overline {CDS} _t^B  = \sum\limits_{i = 1}^{NB} {w_{i,t} CDS_{t_i }^B } }  \\
\end{array}
  \label{eq:rav}
\end{equation}

The conclusions which can be extracted from equation (\ref{eq:rtF}) are the following:
\begin{itemize}
	\item Increasing $CDS_t^B$ implies a profit for the Financial area because it has to pay less recovery to the bond holders. This means that when the credit quality of the bank deteriorates, the internal Fund Transfer Price, FTP, gets reduced due to this profit. 
	\item Assuming that the CDS spread is within the internal average, the FTP rate so that the financial area is not a profit center essentially depends on the average liquidity spread of already issued bonds, $\bar l_{t,s}^B$. This conclusion is also reached by \cite{2017_Acc_vs_Eco_XVA} and \cite{Gunnenson2014}.
\end{itemize}

\begin{equation}
FA_t  = F_t  \cdot \int_t^T {p_{t,s}^r \sum\limits_{i = 1}^{NB} {w_{i,t} \left( {r_{t_i ,s}^{FTP}  - r_{t_i ,s}^B } \right)ds} } 
  \label{eq:FAt_nondefault}
\end{equation}

\begin{equation}
\bar r_{t,s}^B  = \sum\limits_{i = 1}^{NB} {w_{i,t} r_{t_i ,s}^B } 
  \label{eq:rFTP_nondefault}
\end{equation}

Under the hypothesis that the bank cannot default, equation (\ref{eq:FAt_CDS}) turns into equation (\ref{eq:FAt_nondefault}) where the survival probability of the bank, $Q_s^B$ is equal to 1 and the CDS spread disappears as it has been assumed that the default intensity of the bank is zero: $\lambda_{t,s}^B = 0$. For equation (\ref{eq:FAt_nondefault}) to be zero (the Financial area is not a profit center), it implies that the average FTP rate is equal to the average yield of the issued bonds as given by equation (\ref{eq:rFTP_nondefault}). The conclusion is that considering that the bank can default makes the FTP rate dependent on liquidity and if it is assumed that the bank cannot default, the FTP rate is the average yield of issued bonds.

\section{Interaction of CVA, DVA and FVA}
\label{sec:InteractionCVADVAFVA}

This section illustrates that credit, debit and funding value adjustments do not behave independently but are closely related to each other. Having into account this inter-relation clarifies issues which will be reviewed later on in sections \ref{sec:AccountingPerspective} to \ref{sec:ComparisonPerspectives}.

First of all, positions that generate CVA or DVA are related with their collateral requirements. CVA generating positions (those which increase CVA) are positions with positive mark-to-market in favor of the bank (assets) which are uncollateralized, partially collateralized or unilateral where counterparty does not post collateral. Therefore, these positions are assets with positive mark-to-market in favor of the bank which get hedged with collateralized liabilities. The payments of the assets which represent the CVA generating positions cancel the liabilities of the hedges closing out the positions. However, the bank has to post collateral for the hedges as they are collateralized liabilities and this collateral cannot be taken from counterparties (those transactions are uncollateralized). In conclusion, CVA generating positions consume collateral.

DVA generating positions on the other hand are positions with negative mark-to-market in favor of the counterparty (liabilities of the bank) which are uncollateralized, partially collateralized or unilateral where the bank does not have to post collateral. Therefore, these positions are liabilities with negative mark-to-market in favor of the counterparty which get hedged with collateralized assets. The payments of the hedges will cancel the liabilities of the DVA generating positions closing out the positions. However, the bank receives collateral from these hedges (they are collateralized assets) which does not have to post for its liabilities (they are uncollateralized). In conclusion, DVA generating positions provide collateral. The Equity business is an example of this situation. This business sells options to uncollateralized customers (e.g. corporates or private banking) and gets the premium. This premium is invested in hedges to offset the profit and loss and the hedges become assets for the bank. As they are collateralized, the premiums paid to establish the hedges are given back in form of collateral. Therefore, the overall effect of the Equity business is to provide liquidity to the bank in form of collateral. 

\begin{figure}[htbp]
	\centering
	\includegraphics[width=0.9\textwidth]{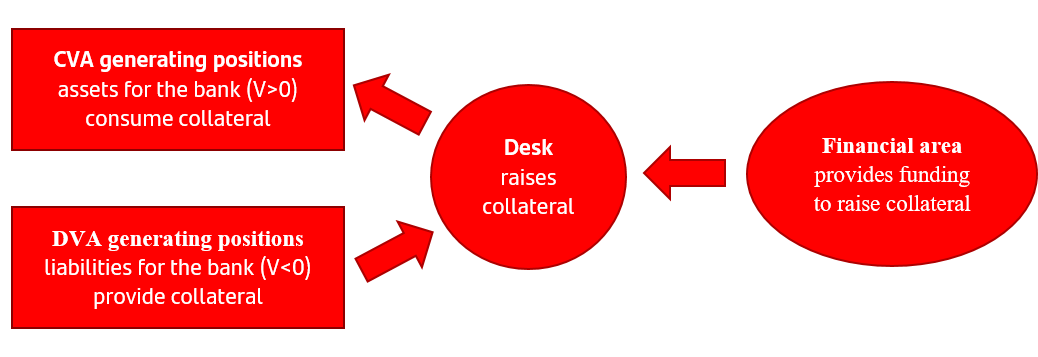}
	\caption{Balance between CVA and DVA generating positions and the Financial area.}
	\label{fig:XVA_balance}
\end{figure}

If re-hypothecation is assumed, the collateral provided by DVA generating positions can be reused for the CVA generating positions by means of the Securities Financing desk. If the collateral consumed by the CVA generating positions is higher than the collateral provided by the DVA generating positions, this collateral must be raised by the desk by borrowing money from the Financial area. This situation is graphically presented in figure \ref{fig:XVA_balance}. If, on the contrary, the amount of collateral provided by DVA generating positions is bigger than the collateral consumed by the CVA generating positions, there is a surplus of collateral which the desk may try to lend to the Financial area (however, this may not always be possible). Therefore, the most optimal management would consist of balancing the amount of collateral provided by DVA generating positions to fulfill the collateral comsumption by CVA generating positions. If this balance is achieved, then neither extra collateral is needed nor there is excess of it. Therefore, it is key to balance CVA and DVA generating positions to minimize collateral requirements.

An imbalance between CVA and DVA generating positions leads to lack or excess of collateral which has to be raised or remunerated producing a financing cost/benefit. However, the following considerations have to be taken into account in relation with this lack/excess of collateral:

\begin{itemize}
	\item Generally speaking, funding benefit obtained from surplus of collateral can only realize the OIS rate (no benefit) unless the collateral excess may reduce bond issuance (this needs a lot of coordination with the Financial area which in practice does not happen) or be reused in repos, initial margin or other instruments through the Securities Financing desk. This increases the return of this collateral but most likely not as much as the internal financing cost which would be achieved by reducing bond issuance.
	\item Funding costs are always realized at the internal bank financing cost which is generally speaking rather expensive. Therefore, this situation should be avoided as much as possible.
	\item CVA/DVA imbalance increases the sensitivity to the financing curve and the need to hedge it. For instance the situation of many banks is an excess of CVA generating positions. These positions consume collateral which has to be raised by the Financial area at a high cost. Therefore, if the funding curve moves up (increase of funding cost), there is a big loss as funding costs get higher.
	\item Increasing CVA rises counterparty debt which reduces internal recovery rate (higher CDS spread). If the CVA/DVA imbalance effectively provides liquidity to the counterparties (uncollateralized) at the expense of the Financial area, the bank increases its counterparty default risk as these counterparties may default.
	\item Increasing DVA rises bank debt leading to potencial systemic risk. From a conceptual point of view, there is nothing wrong by the fact that the bank increases its liabilities if they will be paid back at some point in the future. However, if the bank defaults at some time, those liabilities may produce systemic risk. Therefore, regulators may not be very happy about increasing liabilities of a bank.
\end{itemize}

In conclusion, the best way to manage the collateral produced by CVA and DVA generating positions is to balance them so that neither extra collateral nor surplus is needed.

\section{Invariance principle without default}
\label{sec:InvariancePrinciple}

This section reviews the formulation of the funding invariance principle for which the complete derivation may be found in \cite{2016_InvariancePrinciple}. Consider first the complete set of cash flows $\tilde C_s$ as seen from the desk in equation (\ref{eq:Cs_tilde}).

\begin{equation}
d\tilde C_s^{r*}  = dC_s  + \left( {r_s^*  - r_s^C } \right)M_s ds + \left( {r_s^*  - r_s^F } \right) \hat F_s ds
  \label{eq:Cs_tilde}
\end{equation}

\begin{figure}[htbp]
	\centering
	\includegraphics[width=0.8\textwidth]{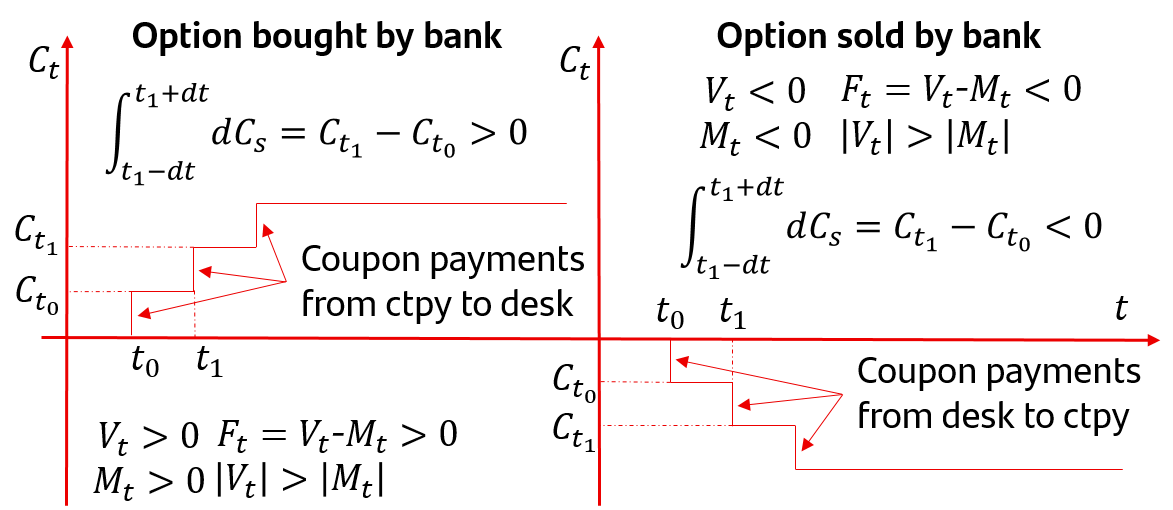}
	\caption{Cash flow process for an option bought (left) and sold (right) by the desk to a counterparty along with sign for derivative price, $V_t$, collateral balance, $M_t$ and funding account, $F_s$.}
	\label{fig:Inv_notation}
\end{figure}

This equation considers a portfolio of derivatives closed with different counterparties which exchange cash flows with the desk. The symbols in this equation mean the following:

\begin{itemize}
	\item $C_s$: cummulative process of contingent cash flows already received ($dC_s>0$) from desk or paid ($dC_s<0$) by desk to the counterparty. These contingent cash flows represent the payments along the life of derivatives. This process has a constant evolution until a cash flow is received or paid by the desk to the counterparty. Figure \ref{fig:Inv_notation} shows how this process looks like. On the left hand side a positive but increasing process indicates cash flows which have been received along the life of the contract (e.g. a vanilla cap bought by the bank) from the counterparty. On the right hand side a decreasing but negative process represents cash flows already paid by the desk (e.g. a vanilla cap sold by the bank) to the counterparty. The integral of this process experiences positive jumps for coupon payments from counterparty to desk and negative ones for payments from desk to counterparty.
	\item $r_s^*$: symmetric and arbitrarily chosen money market rate at which the desk internally borrows/lends money either to counterparties or to the Financial area.
	\item $M_s$: re-hypothecable collateral deposited in desk ($M_s$>0) or posted by desk ($M_s<0$) to counterparties. It gets remunerated at $r_s^C$. Figure \ref{fig:Inv_notation} shows that for options bought by the bank (left hand side), the desk pays a premium and gets a contract, $V_t$, which is positive (the counterparty owes the payoff to the desk) and therefore part of that debt is posted by the counterparty to the desk in form of collateral, $M_t$, which is also positive. For options sold by the bank (right hand side), the desk receives a premium and delivers a contract, $V_t$, which is negative (a liability of the bank) and therefore part of that liability is posted by the desk to the counterparty in form of collateral, $M_t$, which is negative.
	\item $\hat F_s$: funding borrowed ($\hat F_s>0$) or lent ($\hat F_s<0$) by desk from/to Financial area. It gets charged ($\hat F_s>0$) or remunerated ($\hat F_s<0$) the rate $r_s^F$ (possibly asymmetric: borrowing at FTP and lending at OIS). The value of $\hat F_t > 0$ represents the financing not covered by collateral which is lent to the counterparty by the Financial area through the desk. The value of $\hat F_t < 0$ represents the financing not covered by collateral which is lent by the counterparty to the Financial area through the desk. One of the conclusions of the funding invariance principle is that $\hat F_s = \hat V_s - M_s$, where $\hat V_s$ is the risk neutral price of the derivative or netting set under analysis plus the valuation adjustments (collateral, funding, credit and debit).
\end{itemize}

Equation (\ref{eq:Cs_tilde}) states that the increments of the total cummulative cash flow process seen by the desk, $\tilde C_s$, is equal to the increments of the cummulative cash flow process of the derivative, $C_s$, plus the cash flows to fund the collateral received, $M_s >0$ (remunerated at the fictitious rate $r_s^*$ and paying the collateral remuneration rate $r_s^C$), plus the cash flows to fund the liquidity given to counterparties not covered by collateral, $\hat F_s = \hat V_s - M_s > 0$, which is remunerated at the fictitious rate, $r_s^*$, and pays $r_s^F$, the remuneration rate of the Financial area.

\begin{equation}
V_t^{r*}  = {\bf E}_t \left[ {\int_t^T {p_{t,s}^{r*} dC_s } } \right]
  \label{eq:price_Cs}
\end{equation}

\begin{equation}
\begin{array}{l}
 \hat V_t  = {\bf E}_t \left[ {\int_t^T {p_{t,s}^{r*} \left( { dC_s  + ({r_s^*  - r_s^C }) M_s  + ( {r_s^*  - r_s^F } ) \hat F_s } \right)ds} } \right]  \\ 
  \;\;\;\;\;= V_t^{r*}  + {\bf E}_t \left[ {\int_t^T {p_{t,s}^{r*} \left( {( {r_s^*  - r_s^C } ) M_s  + ( {r_s^*  - r_s^F } ) \hat F_s } \right) ds} } \right] \\ 
 \end{array}
  \label{eq:funding_eq}
\end{equation}

\begin{equation}
\hat F_s = \hat V_s - M_s
  \label{eq:cond_principle}
\end{equation}

The funding invariance principle \cite{2016_InvariancePrinciple} is obtained integrating equation (\ref{eq:Cs_tilde}) multiplied by $p_{t,s}^{r*}$, taking conditional expectation (${\bf E}_t \left[  \cdot  \right]$) and replacing equation (\ref{eq:price_Cs}) which relates the risk neutral price discounted with rate $r^*$, with the cummulative process of contingent cash flows. This yields the equation (\ref{eq:funding_eq}) of the invariance principle. This principle states that equation (\ref{eq:funding_eq}) holds irrespective of the choice of $r_s^*$ provided that equation (\ref{eq:cond_principle}) holds, namely, the money borrowed/lent from derivatives not covered by collateral must be taken from the Financial area. This means that the net amount of cash flows coming to the desk from the counterparty, the Financial area and the collateral account is zero. Therefore, their remuneration rate, $r_s^*$, is irrelevant because a zero balance of net cash flows times any rate is still zero.

\section{Invariance principle with default}
\label{sec:InvariancePrincipleDefault}

The invariance funding principle including default risk needs to include the cash flows corresponding to either the counterparty or the bank default. However, it is based on the same idea: the financing provided to counterparties not covered by collateral must come from the Financial area: $\hat F_s = \hat V_s - M_s$. Including credit risk yields equations (\ref{eq:fundcr_eq}) to (\ref{eq:H_XVA}) which are derived in appendix \ref{sec:Invariance_default} (see also p. 20 of \cite{2017_Acc_vs_Eco_XVA}), where $d{\bf 1}_{ \left\{ \tau _C  \le s \right\}}$ is the counterparty default indicator at time $s$ further explained in equation (\ref{eq:dOne}) of appendix \ref{sec:Invariance_default}, $\tilde V_s$ is the exit price on default, $R_s^C$ is the recovery of the counterparty, $\tau_C$ and $\tau_B$ are respectively the times to default of the counterparty and the bank and $\tau = min(\tau_C,\tau_B)$ is the first to default time. See that as long as $\hat F_s = \hat V_s - M_s$ holds, $r_s^*$ can be arbitrarity chosen.

\begin{equation}
\begin{array}{l}
 \hat V_t {\bf 1}_{\left\{ {\tau  > t} \right\}}  = V_t^{r*} {\bf 1}_{\left\{ {\tau  > t} \right\}}  + ColVA_t^{r*}  + FVA_t^{r*}  + CVA_t^{r*}  + DVA_t^{r*}  \\ 
 {\;\;\;\;\;\;\;\;\;\;\;\;\;\;\;\;\;\;\;\;\;\;\;\;\;\;\;\;\;}V_t^{r*}  = {\bf E}_t \left[ {\int_t^\infty  {p_{t,s}^{r*} dC_s } } \right] \\ 
 \end{array}
  \label{eq:fundcr_eq}
\end{equation}

\begin{equation}
\begin{array}{l}
 ColVA_t^{r*}  = {\bf E}_t \left[ {\int_t^\infty  {p_{t,s}^{r*} M_s \left( {r_s^*  - r_s^C } \right){\bf 1}_{ \left\{ \tau  > s \right\} } ds}  } \right] \\ 
 {\;\;}FVA_t^{r*}  = {\bf E}_t \left[ {\int_t^\infty  {p_{t,s}^{r*} \hat F_s \left( {r_s^*  - r_s^F } \right){\bf 1}_{ \left\{ \tau  > s \right\} } ds} } \right] \\ 
 \end{array}
  \label{eq:FColVA}
\end{equation}

\begin{equation}
\begin{array}{l}
 CVA_t^{r*}  = {\bf E}_t \left[ {\int_t^\infty  {p_{t,s}^{r*} \left( {H_s^C  + M_s - V_s^{r*}} \right){\bf 1}_{ \left\{ \tau_B  > s \right\} } d{\bf 1}_{ \left\{ \tau _C  \le s \right\} } } } \right] \\ 
 DVA_t^{r*}  = {\bf E}_t \left[ {\int_t^\infty  {p_{t,s}^{r*} \left( {H_s^B  + M_s - V_s^{r*}} \right){\bf 1}_{ \left\{ \tau_C  > s \right\} } d{\bf 1}_{ \left\{ \tau _B  \le s \right\} } } } \right] \\ 
 \end{array}
  \label{eq:CDVA}
\end{equation}

\begin{equation}
\begin{array}{l}
 H_s^B  = (\tilde V_s  - M_s )^ +   + R_s^B (\tilde V_s  - M_s )^ -   \\ 
 H_s^C  = R_s^C (\tilde V_s  - M_s )^ +   + (\tilde V_s  - M_s )^ -   \\ 
 \end{array}
  \label{eq:H_XVA}
\end{equation}

The Collateral Value Adjustment (ColVA) and funding value adjustment (FVA) terms provided by equation (\ref{eq:FColVA}) come from equation (\ref{eq:funding_eq}). The only change is that the calculation is conditioned by the fact that neither the bank nor the counterparty default (see the indicator function ${\bf 1}_{ \left\{ \tau  > s \right\} }$ inside the integrands).

The Credit Value Adjustment (CVA) and Debit Value Adjustment (DVA) terms provided by equation (\ref{eq:CDVA}) account for the cash flows when either the bank or the counterparty default. The terms $H_s^C$ and $H_s^B$ given by equation (\ref{eq:H_XVA}) and added to the collateral account, $M_s$, represent the default payments of the counterparty and the bank at time $s$. The probability of a joint default (both the counterparty and the bank defaulting simultaneously) is neglected. See that when either the counterparty or the bank defaults, the collateral balance is always part of the default payment (the posted or received collateral is never returned back in case of default).

Equation (\ref{eq:H_XVA}) shows that when the counterparty defaults, it only returns back the recovery fraction of the money it owes to the desk: $R_s^C (\tilde V_s  - M_s )^ +$. See that $\tilde V_s  - M_s$ may be positive for two reasons: either the liquidity provided to the counterparty ($\tilde V_s > 0)$ is not completely covered by the collateral received by the desk ($\tilde V_s > M_s > 0$) or the liability of the bank to the counterparty ($\tilde V_s < 0$) is overcollateralized ($M_s < \tilde V_s < 0$) and from the excess of posted collateral,  only the recovery gets returned. See that the money owed by the desk to the counterparty, $(\tilde V_s  - M_s )^ -$, is always fully returned back when counterparty defaults.

Equation (\ref{eq:H_XVA}) also shows that when the bank defaults, it only returns back the recovery fraction of the money it owes to the counterparty: $R_s^B (\tilde V_s  - M_s )^ -$. Again, $\tilde V_s  - M_s$ may be negative for two reasons: either the liability of the desk to the counterparty, $\tilde V_s < 0$, is not completely covered by the collateral posted by the desk, $\tilde V_s < M_s < 0$, or the money owed by the counterparty to the desk, $\tilde V_s > 0$, is overcollateralized ($M_s > \tilde V_s > 0$) and from the excess of collateral posted to the desk by the counterparty, only the recovery gets returned when the bank defaults.

\begin{equation}
\begin{array}{l}
 CVA_t^{r*}  = {\bf E}_t \left[ {\int_t^\infty  {p_{t,s}^{r*} \left( {H_s^C  + M_s - V_s^{r*}} \right) {\bf E}_s [ {\bf 1}_{ \left\{ \tau_B  > s \right\} } d{\bf 1}_{ \left\{ \tau _C  \le s \right\} } ] } } \right] \\ 
 DVA_t^{r*}  = {\bf E}_t \left[ {\int_t^\infty  {p_{t,s}^{r*} \left( {H_s^B  + M_s - V_s^{r*}} \right) {\bf E}_s [ {\bf 1}_{ \left\{ \tau_C  > s \right\} } d{\bf 1}_{ \left\{ \tau _B  \le s \right\} } ] } } \right] \\ 
 {\;\;}FVA_t^{r*}  = {\bf E}_t \left[ {\int_t^\infty  {p_{t,s}^{r*} \hat F_s \left( {r_s^*  - r_s^F } \right) {\bf E}_s [ {\bf 1}_{ \left\{ \tau  > s \right\} } ] ds} } \right] \\ 
 \end{array}
  \label{eq:CDVA_Es}
\end{equation}

\begin{equation}
\begin{array}{l}
{\bf E}_s [ {{\bf 1}_{\left\{ {\tau _B  > s} \right\}} d{\bf 1}_{\left\{ {\tau _C  \le s} \right\}} } ] \approx {\bf E}_s [ {{\bf 1}_{\left\{ {\tau _B  > s} \right\}} ] {\bf E}_s [ d{\bf 1}_{\left\{ {\tau _C  \le s} \right\}} } ] = e^{ - \int_t^s {( {\lambda _u^C  + \lambda _u^B } )du} } \lambda _s^C ds \\
{\bf E}_s [ {{\bf 1}_{\left\{ {\tau _C  > s} \right\}} d{\bf 1}_{\left\{ {\tau _B  \le s} \right\}} } ] \approx {\bf E}_s [ {{\bf 1}_{\left\{ {\tau _C  > s} \right\}} ] {\bf E}_s [ d{\bf 1}_{\left\{ {\tau _B  \le s} \right\}} } ] = e^{ - \int_t^s {( {\lambda _u^C  + \lambda _u^B } )du} } \lambda _s^B ds \\
\;\;\;\;\;\;\;\;\;\;\;\;\;{\bf E}_s \left[ {{\bf 1}_{\left\{ {\tau > s} \right\}} } \right] \approx {\bf E}_s [{\bf 1}_{\left\{ {\tau_C > s} \right\}}] {\bf E}_s [{\bf 1}_{\left\{ {\tau_B > s} \right\}}] = e^{ - \int_t^s {\left( {\lambda _u^C  + \lambda _u^B } \right)du} } \\
 \end{array}
  \label{eq:d1tauCB}
\end{equation}

Finally, equation (\ref{eq:CDVA}) shows the CVA and DVA terms which are discounted with the arbitrarily chosen rate, $r^*$, multiplying by $p_{t,s}^{r*}$. These payments are weighted by the indicator functions corresponding to the first to default being the counterparty for CVA (${\bf 1}_{\left\{ {\tau _B  > s} \right\}} d{\bf 1}_{\left\{ {\tau _C  \le s} \right\}}$) and the first to default being the bank for DVA (${\bf 1}_{\left\{ {\tau _C  > s} \right\}} d{\bf 1}_{\left\{ {\tau _B  \le s} \right\}}$). An example in which the first-to-default effect may be important can be illustrated by the Regional Communities of Spain. Their default is highly correlated with a default of a Spanish bank. If the Regional Community defaults it is likely that the government of Spain has already defaulted and this would also trigger the default of the Spanish bank.

It is possible to apply the expectation conditioned by time $s$ to the indicator functions of equations (\ref{eq:FColVA}) and (\ref{eq:CDVA}) applying the Tower law as shown by (\ref{eq:CDVA_Es}) as the rest of integrand terms are measurable by time $s$ and can get out of the expectation. Equations (\ref{eq:d1tauCB}) show the expectations of these indicator functions in terms of the hazard rates or default intensities\footnote{Probability of default at time $u$ provided survival prior to time $u$.}, $\lambda_u^B$ and $\lambda_u^C$, of the bank and the counterparty assuming independence between the counterparty and the bank. The contribution of the indicator functions in terms of the harzard rates for the Credit, Debit and Funding value adjustment formulas have also been derived by \cite{Burgard2013} in the context of the resolution of a partial differential equation. The most general case considers that hazard rates and the rest of risk factors are jointly simulated to account for the first-to-default effect and the correlation between market exposure and default. The standard CVA/DVA computation usually considers deterministic hazard rates calibrated at time $t$ . This implies ignoring the correlation between market exposure and default (so called wrong/right way risk) and the first-to-default effect. Some authors have analyzed the impact alone of the first-to-default effect by the use of a copula approach (see \cite{Brigo2011} for more information).

\section{Math framework: objects of selection}
\label{sec:FrameworkObjectsSelection}

This section presents the general framework to consistently calculate credit, debit and funding value adjustments based on the principle of invariance (see \cite{2016_InvariancePrinciple} and \cite{2017_Acc_vs_Eco_XVA}). This framework is partially open to some objects of selection and leads to a family of perspectives which are compatible within the framework. These perspectives focus on how to account and manage these adjustments.

The common framework is defined by equations (\ref{eq:fundcr_eq}) to (\ref{eq:H_XVA}) where different perspectives may be achieved by choosing various objects of selection. Although several perspectives may be included, consistency among them is provided by the reference equations. In particular, one perspective for accounting purposes and another for internal management purposes will be considered. These perspectives depend on the following objects of selection:
\begin{itemize}
	\item \textbf{The way to solve the equation} (\ref{eq:fundcr_eq}): there is a circular relationship as $\hat V_s$, which includes the adjustments, depends on $\hat F_s$ which depends itself on $\hat V_s$. There are various ways to avoid this problem. Sections \ref{sec:NotCircularDep} and \ref{sec:NotCircularDepDefault} address the particular way this circular dependence is avoided for this framework.
	\item \textbf{Curve selection}: there are a number of curves which can be selected: the curve associated with the principle of invariance, $r_s^*$, the collateral remuneration rate, $r_s^C$, and the funding rate charged by the Financial area, $r_s^F$.
	\item \textbf{Exit price}: the selection of $\tilde V_s$ which is assumed to be the price at which the derivatives can be sold to unwind the positions in case of default. This selection may be conditioned by the way the circular dependence in equation (\ref{eq:fundcr_eq}) is solved. The exit price is sometimes calculated under the same valuation assumption as the variation margin (e.g. OIS or risk-free discounting). However, many netting agreements assume that liquidation is carried out at the replacement cost as valued by a third party. This would include the default probability of the surviving party\footnote{A particular case for which there would be risk for the bank in case of counterparty default would be for a position highly in favour of the bank with an unilateral CSA agreement where the counteparty does not post collateral. If the exit price is OIS discounted, the bank would expect to receive more as the replacement cost is always lower than the risk free because the default of the surviving party (the bank) would reduce the price.}.
	\item \textbf{Survival probabilities}: equations (\ref{eq:FColVA}) require bilateral survival of the bank and the counterparty (see factor ${\bf 1}_{\tau  > s}$) and equations (\ref{eq:CDVA}) consider the first-to-default. Therefore, it has to be chosen which of these hypotheses are included in the formulas to calculate the XVA adjustments.
	\item \textbf{Collateral optimisation}: depending on whether the collateral can be re-hypothecated and the ability of reusing collateral by the Securities Financing desk (see section \ref{sec:SecuritiesFinancing}), various options are available to choose the funding rate, $r_s^F$. These choices may reflect accounting or management perspectives.
	
	\begin{itemize}
		\item Assume that funding rates are symmetric (positive and negative balance of collateral gets remunerated/charged the same rate) and common across counterparties. In this situation, the correct calculation considering a single legal entity funding set aggregation level including all the deals of every netting set, see equation (\ref{eq:F_M_LE}) is equivalent to the sum of the funding adjustments at netting set level as expressed in equation (\ref{eq:FVA_ColVA_LE}). The latter approach is the most common among institutions as their systems are not prepared to consider a single funding set for the whole institution instead of considering separate funding sets for each netting set. Adding metrics at netting set level to obtain metrics at legal entity level is always bigger than directly calculating metrics at legal entity level. 
		\item When collateral can be re-hypothecated and reused among all counterparties, it is more reasonable to assume a symmetric funding rate for management purposes under the assumption that the collateral provided by every CSA aggrement is jointly managed by the Securities Financing desk and the funding rate is set to the average rate which considers how excess of collateral is optimized to reduce funding cost. A symmetric rate may also be considered when the bank decides a rate to mark the funding adjustment for accounting purposes (e.g. the liquidity curve shown in section \ref{sec:ExampleFTPrate}).
		\item When collateral is not re-hypothecable or rates are not assumed to be symmetric (e.g. there is not a Securities Financing desk optimizing collateral), then from a management point of view, the aggregation of FVA and ColVA has to be directly aggregated at legal entity level considering a single funding set as indicated by equation (\ref{eq:FVA_ColVA_LE}), where positive collateral balance is remunerated at OIS and negative collateral balance gets charged the internal cost of the bank.
	\end{itemize}
\end{itemize}

These objects of selection share a common framework which is compatible with various uses and perspectives. The bank has to decide which perspectives are most convenient for best internally manage the bank or most satisfy regulation for accounting purposes.

\section{Avoiding circular loop without default\textsuperscript{\ref{fnote:Jerome}}}
\label{sec:NotCircularDep}

This section shows how the circular dependence presented in section \ref{sec:FrameworkObjectsSelection} has been solved for the invariance principle without default. The case with default is described in section \ref{sec:NotCircularDepDefault}.

To solve the circular dependence, the rate associated with the principle of invariance, $r_s^*$, is chosen to be $r_s^F$. This achieves cancelling the term involving $\hat F_s$ in equation (\ref{eq:funding_eq}) and the circular dependence is avoided. However, this is at the cost of having to discount every derivative in the systems with the same rate $r_s^F$: $V_t^{r^F} = {\bf E}_t \left[ \int_t^T {p_{t,s}^{r^F} dC_s}  \right]$ according to equation (\ref{eq:Vhat_VrF}).

\begin{equation}
\hat V_t  = V_t^{r^F} + {\bf E}_t \left[ {\int_t^T {p_{t,s}^{r^F} \left[ {\left( {r_s^F  - r_s^C } \right) M_s } \right]ds} } \right]
  \label{eq:Vhat_VrF}
\end{equation}

As deals are discounted in valuation systems (sys) with another rate, $r^{sys}$, equation (\ref{eq:VrF_Vrmx}) relates $r^F$ and $r^{sys}$ discounting (see section 6.3 of \cite{2014_Acc_FVA_Albanese_Andersen}). This equation has nothing to do with the invariance pricinple. It simply relates two prices discounted with different rates. Appendix \ref{sec:DiscEqDerivation} shows the derivation of this equation.

\begin{equation}
V_t^{r^F }  = V_t^{r^{sys} }  + {\bf E}_t \left[ {\int_t^T {p_{t,s}^{r^F } V_s^{r^{sys} } \left( {r_s^{sys}  - r_s^F } \right)ds} } \right]
  \label{eq:VrF_Vrmx}
\end{equation}

If equation (\ref{eq:VrF_Vrmx}) is replaced in equation (\ref{eq:Vhat_VrF}) and assuming that $F_s^{r^{sys}} = V_s^{r_{sys}} - M_s$, the equation (\ref{eq:Vhat_Vrmx}) is obtained, which avoids the circular dependence and also allows discounting with the rate, $r_s^{sys}$, used for marking purposes within the internal pricing systems:

\begin{equation}
\hat V_t  = V_t^{r^{sys} }  + {\bf E}_t \left[ {\int_t^T {p_{t,s}^{r^F } \left\{ {M_s \left( {r_s^{sys}  - r_s^C } \right) + F_s^{r^{sys} } \left( {r_s^{sys}  - r_s^F } \right)} \right\}} } \right]
  \label{eq:Vhat_Vrmx}
\end{equation}

This equation calculates the collateral and funding value adjustments avoiding the circular dependence and using directly the prices of the derivatives which are already available in the pricing systems.

\section{Avoiding circular loop with default\textsuperscript{\ref{fnote:Jerome}}}
\label{sec:NotCircularDepDefault}

This section presents the solution to avoid the circular loop but considering default risk. This result is presented here under the assumption that the exit price in case of default is assumed to be the price marked in corporate systems: $\tilde V_s = V_s^{r^{sys}}$. Under this hypothesis, the solution is given by equations (\ref{eq:fundcr_Vmx}) to (\ref{eq:CDVA_Vmx}). Appendix \ref{sec:XVAimproved} shows the derivation of these equations from the equations (\ref{eq:fundcr_eq}) to (\ref{eq:H_XVA}) of the general framework presented in section \ref{sec:FrameworkObjectsSelection}.

\begin{equation}
 \hat V_t {\bf 1}_{\left\{ {\tau  > t} \right\}}  = V_t^{r^{sys}} {\bf 1}_{\left\{ {\tau  > t} \right\}}  - ColVA_t^{sys}  - FVA_t^{sys}  - CVA_t^{sys}  - DVA_t^{sys}
  \label{eq:fundcr_Vmx}
\end{equation}

\begin{equation}
\begin{array}{l}
 ColVA_t^{sys}  = {\bf E}_t \left[ {\int_t^\infty  {p_{t,s}^{r^F} M_s \left( {r_s^C - r_s^{sys}} \right){\bf 1}_{\tau  > s} ds} } \right] \\ 
 {\;\;}FVA_t^{sys}  = {\bf E}_t \left[ {\int_t^\infty  {p_{t,s}^{r^F} F_s^{r^{sys}} \left( {r_s^F - r_s^{sys}} \right){\bf 1}_{\tau  > s} ds} } \right] \\ 
 \end{array}
  \label{eq:FColVA_Vmx}
\end{equation}

\begin{equation}
\begin{array}{l}
CVA_t^{sys}  = {\bf E}_t  \left[ {\int_t^\infty  {p_{t,s}^{r^F } \left( {1 - R_s^C } \right)\left( {V_s^{r^{sys} }  - M_s } \right)^ +  {\bf 1}_{\left\{ {\tau_B  > s} \right\}} d{\bf 1}_{\left\{ {\tau _C  \le s} \right\}} } } \right] \\
DVA_t^{sys}  = {\bf E}_t  \left[ {\int_t^\infty  {p_{t,s}^{r^F } \left( {1 - R_s^B } \right)\left( {V_s^{r^{sys} }  - M_s } \right)^ -  {\bf 1}_{\left\{ {\tau_C  > s} \right\}} d{\bf 1}_{\left\{ {\tau _B  \le s} \right\}} } } \right] \\
\end{array}
  \label{eq:CDVA_Vmx}
\end{equation}

See that the adjustments provided by equations (\ref{eq:FColVA_Vmx}) and (\ref{eq:CDVA_Vmx}) must be discounted using the rate chosen for the principle of invariance, $r_s^* = r_s^F$. Now, $F_s^{r^{sys}}$ can be easily calculated as it depends on $V_s^{r^{sys}}$ which is available in the systems. Equation (\ref{eq:CDVA_Vmx}) shows the well known formulas for CVA and DVA depending on the recovery rates of counterparty and bank, $R_s^C$ and $R_s^B$, but with an exit price discounted with the discounting rate used in the corporate systems. This is the major hypothesis of the particular solution which has been chosen to solve equation (\ref{eq:fundcr_eq}).

Equations (\ref{eq:fundcr_Vmx}) to (\ref{eq:CDVA_Vmx}) are defined at deal level. However, the aggregation levels which are considered in practice are CSA, netting set and legal entity. Equation (\ref{eq:Ms_CSA_NS}) shows the variation margin calculated at CSA level for the deals corresponding to the perimeter of the CSA and at netting set level for the CSA contracts included in the netting set.

\begin{equation}
\begin{array}{*{20}c}
   {M_s^{NS_i }  = \sum\limits_{CSA_j  \in NS_i } {M_s^{CSA_j } } } & {} & {M_s^{CSA_j }  = \sum\limits_{Deal_k  \in CSA_j } {M_s^{Deal_k } } }  \\
\end{array}
  \label{eq:Ms_CSA_NS}
\end{equation}

Equation (\ref{eq:F_NS}) shows the mark-to-future position, $V_s^{NS_i}$, at netting set level and the funding position, $F_s^{NS_i}$, in terms of the positions of mark-to-future and variation margin, $M_s^{NS_i}$, at netting set level. See that $V_s^{NS_i}$ includes the deals of every CSA contract and the deals which are outside the CSA agreements of the netting set considered.

\begin{equation}
\begin{array}{*{20}c}
   {F_s^{NS_i }  = V_s^{NS_i }  - M_s^{NS_i } } & {} & {V_s^{NS_i }  = \sum\limits_{Deal_k  \in NS_i } {V_s^{Deal_k } } }  \\
\end{array}
  \label{eq:F_NS}
\end{equation}

Equation (\ref{eq:F_M_LE}) presents the funding position, $F_t^{LE}$, and the variation margin position, $M_s^{LE}$ at legal entity level. They are obtained by summing across netting sets.

\begin{equation}
\begin{array}{*{20}c}
   {F_t^{LE}  = \sum\limits_{NS_i  \in LE} {F_s^{NS_i } } } & {} & {M_s^{LE}  = \sum\limits_{NS_i  \in LE} {M_s^{NS_i } } }  \\
\end{array}
  \label{eq:F_M_LE}
\end{equation}

CVA and DVA are defined at netting set level. Their calculation is carried out using the mark-to-future and variation margin positions at netting set level. Namely, replacing $V_s^{r^{sys}}$ by $V_s^{NS_i}$ and $M_s$ by $M_s^{NS_i}$ in equation (\ref{eq:CDVA_Vmx}).

FVA and ColVA are defined at legal entity level. The calculation is carried out using the funding and variation margin positions at legal entity level. Namely, replacing $F_s^{r^{sys}}$ by $F_s^{LE}$ and $M_s$ by $M_s^{LE}$ in equation (\ref{eq:FColVA_Vmx}).

In practice FVA and ColVA are calculated at netting set level by using the funding and variation margin positions at netting set level. Namely, replacing $F_s^{r^{sys}}$ by $F_s^{NS_i}$ and $M_s$ by $M_s^{NS_i}$ in equations (\ref{eq:FColVA_Vmx}). The legal entity level is thereafter obtained by summing FVA and ColVA across every netting set of the legal entity according to equation (\ref{eq:FVA_ColVA_LE}).

\begin{equation}
\begin{array}{*{20}c}
   {FVA_t^{LE}  = \sum\limits_{NS_i  \in LE} {FVA_s^{NS_i } } } & {} & {ColVA_s^{LE}  = \sum\limits_{NS_i  \in LE} {ColVA_s^{NS_i } } }  \\
\end{array}
  \label{eq:FVA_ColVA_LE}
\end{equation}

See that the aggregation provided by equation (\ref{eq:F_M_LE}) is only true when the rates $r_s^C$ and $r_s^F$ are symmetric (the same irrespective of the sign of $F_s^{r^{sys}}$ or $M_s$). If this is not the case, the correct calculation should be carried out directly at legal entity level. This is something that the systems of most institutions are not prepared for and constitutes an important limitation especially for the calculation of FVA (Funding Value Adjustment) due to the fact that without a Securities Financing desk the excess of collateral ($F_s^{r^{sys}} < 0$) is very likely to earn just the OIS rate rather than $r^F$.

\section{Definition of accounting perspective}
\label{sec:AccountingPerspective}

This section motivates and illustrates the objects of selection described in section \ref{sec:FrameworkObjectsSelection} for an accounting perspective. The accounting perspective is motivated by the current regulation which requires including CVA and DVA and maximize observable inputs (e.g. cost of own funding) for fair value calculation:

\begin{itemize}
		\item Fair value measurement of derivative instruments should include the credit risk of their counterparty as well as their own credit risk (see paragraphs 11, 42, 43 and 56 of \cite{IFRS13} and section 1.1 of \cite{2015_CVACapitalRegulation}).
		\item Methods for fair value measurement should maximise the use of observable inputs and minimise the use of unobservable inputs (paragraph 61 of \cite{IFRS13} and section 1.1 of \cite{2015_CVACapitalRegulation}).
\end{itemize}

In addition, the current regulation requires to de-recognize DVA from calculation of Common Equity Tier 1 capital (CET1): 
\begin{itemize}
		\item The Basel Committee on Banking Supervision establishes with regard to derivative liabilities, to de-recognise all accounting valuation adjustments arising from the bank's own credit risk in the calculation of Common Equity Tier 1 capital (see \cite{BCBS_DVAout} and \cite{DeductionDVAfromCET1}).
\end{itemize}

Given these premises, a proposal for the objects of selection under an accounting perspective could be the following:

\begin{itemize}
	\item \textbf{Solution of equation} (\ref{eq:fundcr_eq}): $r_s^* = r_s^F$ and $\tilde V_s = V_s^{r^{sys}}$ which yields equations (\ref{eq:fundcr_Vmx}) to (\ref{eq:CDVA_Vmx}) as explained in section \ref{sec:InvariancePrincipleDefault}.
	\item \textbf{Curve selection}: $r_s^F = \bar r_{t,s}^{OIS} + \bar l_{t,s}^B$. The implications of this choice are the following:
	\begin{itemize}
		\item The funding rate from the Financial area is set to the Fund Transfer Pricing rate, $r_s^F = \bar r_{t,s}^{FTP}$, according to equation (\ref{eq:rtF}). This selection, as discussed in section \ref{sec:ExampleFTPrate}, is set so that the Financial area pays the recovery of the issued bonds in case of default of the bank and the Financial area is not a profit center.
		\item The average liquidity, $\bar l_{t,s}^B$, of equation (\ref{eq:rtF}) can be replaced by a market proxy (e.g. covered bond, bond-CDS basis of a basket of market representative bonds). Therefore, this liquidity curve would be observable.
		\item The credit default swap (CDS) contribution is also observable and negligible, assuming that there is stability of the CDS. Namely, current CDS is similar to the past average along the bond issuance history.
		\item With this selection there is no overlap between DVA (only considers pure credit curve) and FVA (only consider a pure liquidity curve). This non-overlapping idea is also discussed by \cite{Gunnenson2014}.
	\end{itemize}
	\item \textbf{Exit price}: it is set to the internal marking in systems: $\tilde V_s = V_s^{r^{sys}}$. This choice is conditioned by the solution derived in section \ref{sec:InvariancePrincipleDefault}.
	\item \textbf{Survival probabilities}: unilateral survival for CVA/DVA. This implies that CVA only considers the survival of the counterparty (the indicators ${\bf 1}_{\left\{ {\tau _B  > s} \right\}} d{\bf 1}_{\left\{ {\tau _C  \le s} \right\}}$ of equation (\ref{eq:CDVA_Vmx}) are replaced by $d{\bf 1}_{\left\{ {\tau _C  \le s} \right\}}$) and the DVA only considers the survival of the bank (the indicators ${\bf 1}_{\left\{ {\tau _C  > s} \right\}} d{\bf 1}_{\left\{ {\tau _B  \le s} \right\}}$ of equation (\ref{eq:CDVA_Vmx}) are replaced by $d{\bf 1}_{\left\{ {\tau _B  \le s} \right\}}$). Although this selection is not consistent with the framework (both survival of counterparty and bank should be considered), it complies with accounting regulation (adjustments arising from own credit risk of the bank must be de-recognized from CET1 capital) and it is still bilateral where price agreement may still be achieved if the counterparty also considers unilateral CVA/DVA.
	\item \textbf{Collateral optimisation}: collateral is not optimised. Funding sets are defined at counterparty level with a collateral remuneration rate common for the whole legal entity, symmetric and equal to the liquidity curve given by equation (\ref{eq:rtF}). This choice is observable and equivalent to considering a single funding set for the legal entity and ColVA and FVA metrics would be calculated at netting set level and summed together. The rate $r_s^C$ for the calculation of ColVA would simply consider the remuneration of collateral of each CSA, which is usually symmetric.
\end{itemize}

The main contribution of this perspective is to report XVA metrics comparable among institutions. Therefore, they need to be bilateral so that both the counterparty and the bank may agree on them and they have to use observable market data.

\section{Definition of management perspective}
\label{sec:ManagingPerspective}

This section addresses a perspective of calculating XVA metrics which allows the best practices for the internal management of the bank. These practices include properly fostering portfolio balance between CVA and DVA generating positions (see section \ref{sec:InteractionCVADVAFVA}) and hedging CVA and FVA risk. It is convenient and desirable that the management perspective is also aligned across institutions. Therefore, surveys are useful for aligning best practices. The conclusions according to three surveys, \cite{2017_SurveyFVAasymmetry}, \cite{2018_SurveyXVA} and \cite{Nov2019XVASurveyPwC}, published in 2017, 2018 and 2019, are the following:

\begin{itemize}
	\item Institutions which use internal FTP rates from Financial area to mark FVA similar to equation (\ref{eq:rFTP_nondefault}): 70\% according to figure 10 of \cite{2018_SurveyXVA} and 60\% according to page 3, question Q1 of \cite{Nov2019XVASurveyPwC}.
	\item DVA is not highly used for profit and loss reporting: 82\% as shown in question Q6 of \cite{2017_SurveyFVAasymmetry} and 75\% looking at figure 4, P\&L, of \cite{2018_SurveyXVA}.
	\item DVA is reported in accounting statements: 60\% according to question Q7 of \cite{2017_SurveyFVAasymmetry} and 70\% as explained in figure. 4, P\&L, of \cite{2018_SurveyXVA}.
	\item Institutions which use symmetric FVA: 82\% according to question Q6 of \cite{2017_SurveyFVAasymmetry} and 80\% as shown in figure 4, P\&L, and figure 11 of \cite{2018_SurveyXVA}).
	\item 60\% of funding sets are considered at counterparty level and 32\% at whole bank level as shown in question Q8 of \cite{2017_SurveyFVAasymmetry})\footnote{Both approaches are equivalent only if funding rate is symmetric. The sum of FVA for every counterparty is equal to the FVA of a single funding set whose perimeter  of deals is the union of the perimeters for all counterparties}.
	\item Institutions which hedge FVA: 50\% according to figure 10 of \cite{2018_SurveyXVA} and 64\% according to question Q21 of \cite{Nov2019XVASurveyPwC}.
	\item Institutions which hedge FVA internally against the Financial area: only 35\% according to figure 10 of \cite{2018_SurveyXVA} and 54\% according to question Q21 of \cite{Nov2019XVASurveyPwC}.
\end{itemize}

According to these surveys, market consensus is clearly trending to the use of internal FTP rates, not including DVA in P\&L formula, marking with symmetric funding rates (same rate for borrowing and lending) and hedging FVA against the financial area. Institutions still consider funding sets at counterparty level instead of the legal entity level. Marking with symmetric rates is most likely a convention rather than a fact (see \cite{Gregory_asymmetric_funding}) as monetizing funding benefit is only possible either by reducing bond issuance (term funding against Financial area) or reusing excess of collateral through the Securities Financing desk (see section \ref{sec:InteractionCVADVAFVA}) for other activities in the bank such as posting initial margin, novate deals to more asset heavy, etc.

Given this information from market consensus, the management perspective assumes that the bank is itself default free. This hypothesis is also assumed by FRTB-CVA regulation (see discussion of section \ref{sec:ComparisonRegulation}). This fact aligns the reduction of profit and loss risk due to XVA with the reduction of CVA regulatory capital. A possible proposal of a management perspective assuming that the bank cannot default is the following:

\begin{itemize}
	\item \textbf{Solution of equation} (\ref{eq:fundcr_eq}): $r_s^* = r_s^F$ and $\tilde V_s = V_s^{r^{sys}}$ which yields equations (\ref{eq:fundcr_Vmx}) to (\ref{eq:CDVA_Vmx}) as explained in section \ref{sec:InvariancePrincipleDefault}. This is the same solution adopted for the accounting perspective.
	\item \textbf{Curve selection}: $r_s^F  = \bar r_{t,s}^B  = \sum\limits_{i = 1}^{NB} {w_{i,t} r_{t_i ,s}^B }$. The implications of this choice are the following:
	\begin{itemize}
		\item The funding rate from the Financial area is set to the average yield of issued bonds, $r_s^F = \bar r_s^B$, according to equation (\ref{eq:rFTP_nondefault}). This selection, as discussed in section \ref{sec:ExampleFTPrate}, assumes that the bank cannot default. In practice $r_s^F$ will be set to an average rate which takes into account the FTP internal rates given by (\ref{eq:rFTP_nondefault}) and the collateral optimisation carried out by the Securities Financing desk (excess of collateral may get more remuneration than just the overnight OIS rate).
		\item With this selection there is also no overlap between DVA and FVA. DVA does not exist as bank default is not considered and the FVA includes both the liquidity and credit components as the average yield of the issued bonds includes them both.
	\end{itemize}
	\item \textbf{Exit price}: it is set to the internal marking in systems: $\tilde V_s = V_s^{r^{sys}}$. This choice is conditioned, similarly to the accounting perspective, by the solution derived in section \ref{sec:InvariancePrincipleDefault}.
	\item \textbf{Survival probabilities}: as DVA disappears because bank default is not considered, CVA turns naturally unilateral and FVA is discounted by $r_s^F = \bar r_s^B$, an average bond issuance cost curve.
	\item \textbf{Collateral optimisation}: as already mentioned before, the Securities Financing desk optimizes re-hypothecable collateral to reuse it up for instance 85\% and determines an average symmetric rate $r_s^F$ which takes into account the optimization of collateral as a whole. With this common symmetric rate, calculating funding sets for the legal entity or for each counterparty netting set would be equivalent (the total FVA would be the sum of FVA across every counterparty). The rate $r_s^C$ for the calculation of ColVA would simply consider the remuneration of collateral of each CSA, which is symmetric.
\end{itemize}

The main contribution of this perspective is to foster appropriate management of XVA by allowing portfolio balance between CVA and DVA generating positions and hedging of CVA and FVA market risk as well as counterparty default risk from CVA.

\section{Comparison among both perspectives}
\label{sec:ComparisonPerspectives}

This section compares the advantages and disadvantages of both accounting and management perspectives. They have different uses and therefore they address complementary parts of the XVA managing and reporting. The advantages and disadvantages of the accounting perspective are the following:

\begin{itemize}
	\item The XVA metrics and the fair value of derivative prices including adjustments are comparable among institutions as adjustments are bilateral (include information of the both the bank and the counterparty) and they are calculated with market observable data.
	\item Including CVA and DVA in profit and loss allows fostering balancing CVA and DVA generating positions hedging one with the other.
	\item The main disadvantage of this perspective is that hedging counterparty default and market funding risk is not possible:
	\begin{itemize}
		\item Hedging CVA default imbalances CVA/DVA compensation. Therefore, the use of credit hedging products such as credit default swaps, pass through swaps or risk participation agreements will not allow offsetting risks.
		\item The liquidity curve is up to some extent a theoretical construct which is good for marking purposes. However, the exposure of the metrics to the movement of this curve can be very material and hedging it may not make a lot of sense.
		\item DVA cannot be hedged in practice.
	\end{itemize}
\end{itemize}

The advantages and disadvantages of the management perspective are the following:
\begin{itemize}
	\item The main disadvantage of this approach is that fair value and XVA metrics are not comparable among banks, as metrics are not bilateral (DVA is absent) and depend on an unobservable internal funding cost (the average price of bonds across the issuing history of the bank optimized by the Securities Financing desk).
	\item CVA credit and market risk can be hedged as there is no DVA to compensate CVA risk.
	\item FVA can be hedged with internal term operations between desk and Financial area.
\end{itemize}

The fact that the management perspective does not provide comparable prices among institutions (law of one price is broken) should not be a major problem for the following reasons:
\begin{itemize}
	\item There should not be great price discrepancies for operations between big institutions as they will be fully collateralized and therefore the size of the credit, debit and funding value adjustments will be small.
	\item The cases for which bigger discrepancies may appear correspond to unsecured operations of usually non-financial institutions which will accept the price provided by the other counterparty as they cannot either effectively challenge the price or have access to an alternative counterparty.
\end{itemize}

\section{Comparison with FRTB-CVA regulation}
\label{sec:ComparisonRegulation}

The regulation establishes that market risk uncertainty should be covered by an amount of capital currently given by the value-at-risk metric. This framework will be replaced in the future by the FRTB or Fundamental Review of the Trading Book (see \cite{FRTB2019}) which calculates the expected shortfall instead of the value-at-risk. For CVA, a separate value-at-risk calculation is currently carried out and will be also replaced in January 2022 by what is called the FRTB-CVA regulation (see \cite{Basel3_CVA}). The main objectives of this new regulation are the following:

\begin{itemize}
	\item Capture both credit and market risk factors when calculating regulatory CVA capital and recognize market risk hedging to reduce it.
	\item Align the regulatory CVA calculation formula with the fair value CVA which is incorporated in the profit and loss (P\&L) as shown in equation (\ref{eq:CDVA_Vmx}). This means that the CVA formula used to calculate capital considers the same risk neutral valuation formula as the CVA calculation for P\&L.
	\item Align regulatory CVA with FRTB calculation rules (see \cite{FRTB2019}). This implies that the expected shortfall of the CVA is calculated through a parametric method based on the sensitivities of the CVA with respect to market and credit risk factors.
\end{itemize}

The analysis of the regulation concludes that the standard approach for CVA calculation (FRTB-CVA) is well aligned with the management perspective for the following reasons:
\begin{itemize}
	\item Regulatory CVA excludes bank own default from regulatory CVA calculation (see paragraph 1 of \cite{Basel3_CVA}): this implies unilateral CVA without DVA.
	\item Regulatory CVA is calculated with the same risk free implied market calibration as P\&L CVA (see paragraphs 31-34 of \cite{Basel3_CVA}).
	\item It recognizes market \& credit hedges: reducing P\&L risk implies reducing CVA capital (see paragraph 37 of \cite{Basel3_CVA}).
\end{itemize}

The conclusion is that the regulation is pointing to the management perspective. In this context, the most sensible approach would be to align accounting regulation with the management perspective so that the profit and loss for internal management is the one reported for accounting purposes (this is not currently the case as accounting regulation requires reporting of CVA and DVA). However, in order to allow regulators to compare figures among institutions, reporting of CVA and DVA figures calculated according to the accounting perspective may be additionally required.

\section{Transition from accounting to management}
\label{sec:Transition}

According to sections \ref{sec:ManagingPerspective} and \ref{sec:ComparisonRegulation}, the market consensus and the regulatory trends are pointing to the management perspective in the future. The implications of changing from the P\&L formula (\ref{eq:EQAccount}) of the accounting perspective to the P\&L formula (\ref{eq:EQManage}) of the management perspective are the following:

\begin{itemize}
	\item The DVA profit is eliminated from the P\&L (a loss must be realized).
	\item Funding cost increases as the average bank issuance yield ($r^F = \bar r^B$) given by equation (\ref{eq:rFTP_nondefault}) is higher than the average bond liquidity spread ($r^F = \bar r^{OIS}  + \bar l^B$) given by equation (\ref{eq:rtF}). This implies that a bank with a net asset position\footnote{Counterparties owe money to the bank and bank must post collateral to hedge.} will have to realize a loss. On the other hand, a bank with a net liability position may realize a profit, if benefit on collateral excess can be monetized.
\end{itemize}

\begin{equation}
\hat V_t  = V_t^{r^{sys} }  - ColVA_t^{\bar r^{OIS}  + \bar l^B }  - FVA_t^{\bar r^{OIS}  + \bar l^B }  - CVA_t^{\bar r^{OIS}  + \bar l^B }  - DVA_t^{\bar r^{OIS}  + \bar l^B }
  \label{eq:EQAccount}
\end{equation}

\begin{equation}
\hat V_t  = V_t^{r^{sys} }  - ColVA_t^{\bar r^B }  - FVA_t^{\bar r^B }  - CVA_t^{\bar r^B } 
  \label{eq:EQManage}
\end{equation}

To implement the transition without negatively affecting shareholders, benefits on new deals should be provisioned and the perspective switched from the accounting formula (\ref{eq:EQAccount}) to the management formula (\ref{eq:EQManage}) when this provision exceeds the loss of the change of perspective. New deals should be marked to funding curves which match market prices. For institutions with heavy asset portfolios the transition implies a loss and marking positions to a market transitioning to the management perspective will also reduce benefits (as compared with a marking according to the accounting perspective). This implies that in order to raise the transition loss, the provisioning period will be longer with the associated business undermining.



\section{Conclusions}
\label{sec:Conclusions}

Two paradigms or perspectives have been defined for consistent credit, debit and funding value adjustment calculation which share the same mathematical framework and implementation but change the input parameters. The accounting perspective considers that CVA, DVA and FVA, are aligned with current accounting regulation and provides symmetric prices based on observable market parameters comparable among institutions. The management perspective only considers CVA and FVA, allows hedging of market risk factors and counterparty default but does not provide adjustments comparable among institutions. Market consensus trends and FRTB-CVA regulation for CVA capital calculation are pointing towards the management perspective. However, there are still a considerable amount of participants which are either in an accounting perspective or in the transition process to a management perspective. For many institutions which have asset heavy portfolios (their counterparties owe money to them), switching to a management perspective implies a loss which justifies the resistance to complete the transition until the accounting regulation and market consensus get more clearly clarified.

The mathematical framework presented in this paper provides an improved solution for the equations of the principle of invariance in which it is based. This allows the calculation of credit, debit and funding adjustment metrics eliminating some approximations of the original paper \cite{2016_InvariancePrinciple} and in terms of the exposures already available in the pricing systems irrespective of the discounting curve chosen for their valuation.

The paper describes how the Financial area works and how balance sheet financing is carried out. It also provides an illustration of how funding rates can be estimated under the assumption that the Financial area is not a profit center. For the accounting perspective in which it is assumed that the bank can default, the conclusion is that the internal funding rate is approximately equal to the average OIS rate plus the liquidity spread of the issued bonds of the institution weighted by their issuance size. On the other hand, for the management perspective in which it is assumed that the bank cannot default, the internal funding rate is the optimized management rate based on the internal FTP (Fund Transfer Pricing) rate and the funding savings out of the collateral optimisation.

\section{Acknowledgments}
\label{sec:Acknow}

The author of this paper wants to acknowledge the important contribution of Jérôme Maetz who provided the improved solution which avoids the circular dependence (see sections \ref{sec:NotCircularDep} and \ref{sec:NotCircularDepDefault}) and very enriching discussions throughout the development process of this paper. Thanks to Carlos Catalán who provided ideas for the specification of the common framework and how to configure it for switching perspectives (section \ref{sec:InvariancePrincipleDefault}). There were key discussions that the author wants to thank to Manuel Villa about collateral optimisation and to Robert Smith, Steven Brittan, Gerard Morris and Enrique Rigol from the XVA desk. The author found a very collaborative atmosphere which allowed collecting the descriptive material about balance sheet management of section \ref{sec:BalanceSheet}. Thank you very much to Mercedes Mora and Carmen del Pozo (Financial Management Control), Andrés Castro and Antonio Torío (Financial area), Carmen Lafont (Short Term desk) and Juan Manuel Bravo, Enrique Verdú and Héctor Ciruelos (Securities Financing desk). Finally, the author wants to thank Manuel Menéndez, my direct supervisor, for supporting this research. Without the contribution of these people, writing this paper would have not been possible. Thank you very much to them all.

\appendix

\section{Relation between two discounted prices}
\label{sec:DiscEqDerivation}

This section shows the derivation of equation (\ref{eq:VrF_Vrmx}) which relates two prices discounted with two different curves (in this case $r^F$ and $r^{sys}$) through an adjustment.

\begin{equation}
d\left( { p_{t,s}^r V_s^r } {\bf 1}_{\left\{ {\tau  > s} \right\}} \right) = p_{t,s}^r \left( {dV_s  - r_s V_s^r ds} \right) {\bf 1}_{\left\{ {\tau  > s}  \right\}}  + p_{t,s}^r V_s^r d{\bf 1}_{\left\{ {\tau > s} \right\}}
  \label{eq:dexpV}
\end{equation}

The left hand side of equation (\ref{eq:dexpV}) shows the differential of the product of the exponential term, $p_{t,s}^r$, defined in (\ref{eq:Pts_Qts}), the stochastic price, ${V_s}^r$, discounted by a generic curve $r$ and the survival indicator function, ${\bf 1}_{\left\{ {\tau  > s} \right\}}$, where $\tau$ is the time of the first to default. Applying Ito's Lemma to this product (second derivatives are zero) and the fundamental theorem of calculus to the exponent of the exponential yields the right hand side of equation (\ref{eq:dexpV}).

\begin{equation}
\int_t^\infty  {d\left( {p_{t,s}^r V_s^r } {\bf 1}_{\left\{ {\tau  > s} \right\}}  \right)}  =  \left. {p_{t,s}^r V_s^r {\bf 1}_{\left\{ {\tau  > s} \right\}} } \right|_{s \to t}^{s \to \infty } =  - V_t^r {\bf 1}_{\left\{ {\tau  > t} \right\}}
  \label{eq:Int_dexpV}
\end{equation}

\begin{equation}
\begin{array}{l}
\int_t^\infty  {d\left( {p_{t,s}^r V_s^r {\bf 1}_{\left\{ {\tau  > s} \right\}} } \right)}  = \int_t^\infty  {p_{t,s}^r  { (dV_s  - r_s V_s^r ds) {\bf 1}_{\left\{ {\tau  > s} \right\}} - p_{t,s}^r V_s^r d{\bf 1}_{\left\{ {\tau \le s} \right\}}} } \\
\;\;\;\;\;\;\;\;\;\;\;\;\;\;\;\;\;\;\;\;\;\;\;\;\;\;\;\;\;\;\; = - V_t^r {\bf 1}_{\left\{ {\tau  > t} \right\}}
\end{array}
  \label{eq:Int_dexpV2}
\end{equation}

\begin{equation}
d{\bf 1}_{\left\{ {\tau  > s} \right\}} = {\bf 1}_{\left\{ {\tau  > s + ds} \right\}}  - {\bf 1}_{\left\{ {\tau  > s} \right\}}  =  {\bf 1}_{\left\{ {s < \tau  \le s + ds} \right\}}  =  -d{\bf 1}_{\left\{ {\tau  \le s} \right\}} 
  \label{eq:dOne_tau_g_s}
\end{equation}

Integrating the left hand side of equation (\ref{eq:dexpV}) by applying Barrow's rule yields equation (\ref{eq:Int_dexpV}). Equation (\ref{eq:Int_dexpV2}) arises from (\ref{eq:dexpV}) and (\ref{eq:Int_dexpV}) after replacing $d{\bf 1}_{\left\{ {\tau  > s} \right\}} = -d{\bf 1}_{\left\{ {\tau \le s} \right\}}$ according to equation (\ref{eq:dOne_tau_g_s}). If conditional expectations are taken on equation (\ref{eq:Int_dexpV2}), equation (\ref{eq:E_dexpV}) is obtained. This equation will be used later in this section and in appendix \ref{sec:Invariance_default} to integrate the price conditioned by the survival of the bank and the counterparty.

\begin{equation}
{\bf E}_t \left[ {\int_t^\infty {p_{t,s}^r { (dV_s^r  - r_s V_s^r ds) {\bf 1}_{\left\{ {\tau  > s} \right\}} - p_{t,s}^r V_s^r d{\bf 1}_{\left\{ {\tau \le s} \right\}} } }  } \right] =  - V_t^r {\bf 1}_{\left\{ {\tau  > t} \right\}}
  \label{eq:E_dexpV}
\end{equation}

Replacing equation (\ref{eq:price_Cs}) with $r^* = r$ in equation (\ref{eq:E_dexpV}), taking the indicator function inside the expectation as it is measurable at time $t$ and reorganizing yields equation (\ref{eq:E_dexpV_Cs}). See that the indicator function multiplying $dC_s$ depends on $t$ and not on $s$ as the other indicator functions of the integral.

\begin{equation}
{\bf E}_t \left[ {\int_t^\infty {p_{t,s}^r { (dV_s^r  - r_s V_s^r ds) {\bf 1}_{\left\{ {\tau  > s} \right\}} - p_{t,s}^r V_s^r d{\bf 1}_{\left\{ {\tau \le s} \right\}} + {\bf 1}_{\left\{ {\tau  > t} \right\}} dC_s } }  } \right] = 0
  \label{eq:E_dexpV_Cs}
\end{equation}

To obtain the relation between two prices discounted with different rates, the rate $r$ of equation (\ref{eq:E_dexpV_Cs}) is set to the two rates, $r_s^F$ and $r_s^{sys}$, to relate with each other and equation (\ref{eq:E_dexpV_Cs}) is divided by $p_{t,s}^r$. This yields equations (\ref{eq:E_dV_rV_F}) and (\ref{eq:E_dV_rV_sys}). See that in the latter equation, $r_s^F$ is added and subtracted to the parenthesis of the second term of the integrand with no effect.

\begin{equation}
{\bf E}_t \left[ {\int_t^\infty  \begin{array}{l}
  (dV_s^{r^F }  - r_s^F V_s^{r^F } ds){\bf 1}_{\left\{ {\tau  > s} \right\}}  \\ 
  - V_s^{r^F } d{\bf 1}_{\left\{ {\tau  \le s} \right\}}  + {\bf 1}_{\left\{ {\tau  > t} \right\}} dC_s  \\ 
 \end{array} } \right] = 0
  \label{eq:E_dV_rV_F}
\end{equation}

\begin{equation}
{\bf E}_t \left[ {\int_t^\infty  \begin{array}{l}
  \left( {dV_s^{r^{sys} }  - (r_s^{sys}  + r_s^F  - r_s^F )V_s^{r^{sys} } ds} \right){\bf 1}_{\left\{ {\tau  > s} \right\}}  \\ 
  - V_s^{r^{sys} } d{\bf 1}_{\left\{ {\tau  \le s} \right\}}  + {\bf 1}_{\left\{ {\tau  > t} \right\}} dC_s  \\ 
 \end{array} } \right] = 0
  \label{eq:E_dV_rV_sys}
\end{equation}

Now equation (\ref{eq:E_dV_rV_sys}) is multiplied by $-1$ and added to equation (\ref{eq:E_dV_rV_F}) yielding equation (\ref{eq:rF_minus_rsys}). The terms multiplying $dC_s$ cancel with each other.

\begin{equation}
{\bf E}_t \left[ {\int_t^\infty  \begin{array}{l}
 \left( {d(V_s^{r^F }  - V_s^{r^{sys} } ) - r_s^F (V_s^{r^F }  - V_s^{r^{sys} } )ds} \right){\bf 1}_{\left\{ {\tau  > s} \right\}}  \\ 
  - (V_s^{r^F }  - V_s^{r^{sys} } )d{\bf 1}_{\left\{ {\tau  \le s} \right\}}  + (r_s^{sys}  - r_s^F )V_s^{r^{sys} } {\bf 1}_{\left\{ {\tau  > s} \right\}} ds \\ 
 \end{array} } \right] = 0
  \label{eq:rF_minus_rsys}
\end{equation}

If equation (\ref{eq:rF_minus_rsys}) is multiplied by $p_{t,s}^{r^F}$, the terms multiplying ${\bf 1}_{\left\{ {\tau  > s} \right\}}$ and $d {\bf 1}_{\left\{ {\tau \le s} \right\}}$ can be solved using equation (\ref{eq:E_dexpV}) with $V_s^r = V_s^{r^F} - V_s^{r^{sys}}$ yielding equation (\ref{eq:VrF_Vrsys}).

\begin{equation}
0 =  - \left( {V_t^{r^F }  - V_t^{r^{sys} } } \right){\bf 1}_{\left\{ {\tau  > t} \right\}}  + {\bf E}_t \left[ {\int_t^\infty  {p_{t,s}^{r^F } (r_s^{sys}  - r_s^F )V_s^{r^{sys} } {\bf 1}_{\left\{ {\tau  > s} \right\}} ds} } \right]
  \label{eq:VrF_Vrsys}
\end{equation}

Reorganizing terms yields equation (\ref{eq:VrF_Vrsys_def}). This is the expression to be proved which relates two prices discounted with different curves.

\begin{equation}
V_t^{r^F } {\bf 1}_{\left\{ {\tau  > t} \right\}} = V_t^{r^{sys} } {\bf 1}_{\left\{ {\tau  > t} \right\}}  + {\bf E}_t \left[ {\int_t^\infty  {p_{t,s}^{r^F } (r_s^{sys}  - r_s^F )V_s^{r^{sys} } {\bf 1}_{\left\{ {\tau  > s} \right\}} ds} } \right]
  \label{eq:VrF_Vrsys_def}
\end{equation}

\begin{equation}
V_t^{r^F }  = V_t^{r^{sys} }  + {\bf E}_t \left[ {\int_t^\infty  {p_{t,s}^{r^F } (r_s^{sys}  - r_s^F )V_s^{r^{sys} } ds} } \right]
  \label{eq:VrF_Vrsys_non_def}
\end{equation}

If it is assumed that there is no default, $\tau \to \infty$, then the indicator functions are always equal to one and equation (\ref{eq:VrF_Vrsys_def}) turns into equation (\ref{eq:VrF_Vrsys_non_def}).

\section{Derivation of invariance with default risk}
\label{sec:Invariance_default}

This section shows the derivation of equations (\ref{eq:fundcr_eq}) to (\ref{eq:CDVA}) of the mathematical framework common to both perspectives (see sections \ref{sec:AccountingPerspective} and \ref{sec:ManagingPerspective}). The starting point is the financing condition, $\xi_s$, of equation (\ref{eq:FCs}) which must be satisfied at any time for the principle of invariance to hold. See section \ref{sec:InvariancePrinciple} for the definition of these symbols. 

\begin{equation}
\xi _s  = \hat V_s  - M_s  - \hat F_s  = 0
  \label{eq:FCs}
\end{equation}

The evolution of the financing condition from time $s$ to $s+ds$ may consider three excluding scenarios: $\xi_{s+ds}^S$ when both the bank and the counterparty survive, $\xi_{s+ds}^C$ when the counterparty defaults but the bank survives, $\xi_{s+ds}^B$ when the bank defaults but the counterparty survives and $\xi_{s+ds}^{BC}$ when both simultaneously default. Equation (\ref{eq:FCsds2}) shows the financing condition at time $s+ds$ considering these scenarios discriminated by indicator functions. See that the indicator functions are expressed in terms of the first to default time $\tau = min(\tau_C,\tau_B)$, where $\tau_C$ and $\tau_B$ are the times to default of respectively the counterparty and the bank.

\begin{equation}
\begin{array}{l}
 \xi _{s + ds}^S {\bf 1}_{\left\{ {\tau  > s + ds} \right\}}  + \xi _{s + ds}^C {\bf 1}_{\left\{ {s < \tau  \le s + ds,\tau  = \tau _C } \right\}}  + \xi _{s + ds}^B {\bf 1}_{\left\{ {s < \tau  \le s + ds,\tau  = \tau _B } \right\}}  \\ 
\;\;\;\;\;\;\;\;\;\;\;\;\;\;\;\;\;\;\;\;\;\;  + \xi _{s + ds}^{BC} {\bf 1}_{\left\{ {s < \tau  \le s + ds,\tau  = \tau _B ,\tau  = \tau _C } \right\}}  = 0 \\ 
 \end{array}
  \label{eq:FCsds2}
\end{equation}

\begin{equation}
\begin{array}{l}
 {\bf 1}_{\left\{ {s < \tau  \le s + ds,\tau  = \tau _C } \right\}}  = {\bf 1}_{\left\{ {\tau_B  > s} \right\}} {\bf 1}_{\left\{ {s < \tau _C  \le s + ds} \right\}} = {\bf 1}_{\left\{ {\tau_B  > s} \right\}} d{\bf 1}_{\left\{ {\tau _C  \le s} \right\}}  \\ 
 {\bf 1}_{\left\{ {s < \tau  \le s + ds,\tau  = \tau _B } \right\}}  = {\bf 1}_{\left\{ {\tau_C  > s} \right\}} {\bf 1}_{\left\{ {s < \tau _B  \le s + ds} \right\}} = {\bf 1}_{\left\{ {\tau_C  > s} \right\}} d{\bf 1}_{\left\{ {\tau _B  \le s } \right\}} \\ 
 \end{array}
  \label{eq:OneFCsds}
\end{equation}

Under the assumption that simultaneous defaults cannot happen (they happen one after another), the last term of equation (\ref{eq:FCsds2}) can be neglected and the indicator functions can be simplified according to equation (\ref{eq:OneFCsds}).

\begin{equation}
\xi _{s + ds}^S {\bf 1}_{\left\{ {\tau  > s + ds} \right\}}  + \xi _{s + ds}^C {\bf 1}_{\left\{ {\tau_B  > s} \right\}} {\bf 1}_{\left\{ {s < \tau _C  \le s + ds} \right\}} + \xi _{s + ds}^B {\bf 1}_{\left\{ {\tau_C > s} \right\}} {\bf 1}_{\left\{ {s < \tau _B  \le s + ds} \right\}} = 0
  \label{eq:FCsds}
\end{equation}

\begin{equation}
\begin{array}{l}
 \xi _{s + ds}^S  = \hat V_{s + ds}  + dC_s  - M_s (1 + r_s^C ds) - \hat F_s (1 + r_s^F ds) \\
 \xi _{s + ds}^C  = H_{s + ds}^C  - \hat F_s (1 + r_s^F ds) \\ 
 \xi _{s + ds}^B  = H_{s + ds}^B  - \hat F_s (1 + r_s^F ds) \\ 
 \end{array}
  \label{eq:xiSCB}
\end{equation}

The value of the financing condition at time $s+ds$ for the three scenarios is shown in equation (\ref{eq:FCsds}) where $\xi _{s + ds}^S$, $\xi _{s + ds}^C$ and $\xi _{s + ds}^B$ are given by equation (\ref{eq:xiSCB}). For the first scenario, the position on derivatives, $\hat V$, takes the value at $s+ds$ plus the payments in the interval $s$ to $s+ds$ provided by $dC_s$. The term, $M_s$, associated with the collateral and the term, $\hat F_s$, associated with the financing, simply accrue the corresponding interest along the interval. For the other two scenarios, the terms $H_{s+ds}^C$ and $H_{s+ds}^B$ are given by equation (\ref{eq:HCB_sds}) and account for the position on derivatives minus the collateral amount at time $s+ds$ considering the default events. See that equation (\ref{eq:HCB_sds}) corresponds to equation (\ref{eq:H_XVA}) but evaluated at time $s+ds$ with the already mentioned considerations of the evolution of derivatives and collateral through the period from time $s$ to $s+ds$.

\begin{equation}
\begin{array}{l}
 H_{s + ds}^C  = R_s^C \left( {\tilde V_{s + ds}  + dC_s  - M_s (1 + r_s^C ds)} \right)^ +   +  \\ 
 \;\;\;\;\;\;\;\;\;\;\;\;\; \left( {\tilde V_{s + ds}  + dC_s  - M_s (1 + r_s^C ds)} \right)^ -   \\ 
 H_{s + ds}^B  = \left( {\tilde V_{s + ds}  + dC_s  - M_s (1 + r_s^C ds)} \right)^ +   +  \\ 
  \;\;\;\;\;\;\;\;\;\;\;\;\;R_s^B \left( {\tilde V_{s + ds}  + dC_s  - M_s (1 + r_s^C ds)} \right)^ -   \\ 
 \end{array}
  \label{eq:HCB_sds}
\end{equation}

Equation (\ref{eq:dOne}) transforms the indicator functions of equation (\ref{eq:FCsds}) in terms of the differential of the indicator function (the unconditional default condition).

\begin{equation}
\begin{array}{l}
 {\bf 1}_{\left\{ {s < \tau _C  \le s + ds} \right\}}  = {\bf 1}_{\left\{ {\tau _C  \le s+ds} \right\}} - {\bf 1}_{\left\{ {\tau _C  \le s} \right\}} = d{\bf 1}_{\left\{ {\tau _C  \le s} \right\}}  \\ 
 {\bf 1}_{\left\{ {s < \tau _B  \le s + ds} \right\}}  = {\bf 1}_{\left\{ {\tau _B  \le s+ds} \right\}} - {\bf 1}_{\left\{ {\tau _B  \le s} \right\}} = d{\bf 1}_{\left\{ {\tau _B  \le s} \right\}}  \\ 
 {\bf 1}_{\left\{ {\tau  > s + ds} \right\}}  = {\bf 1}_{\left\{ {\tau  > s} \right\}} - {\bf 1}_{\left\{ {s < \tau \le s + ds} \right\}} = {\bf 1}_{\left\{ {\tau  > s} \right\}}  - d{\bf 1}_{\left\{ {\tau  \le s} \right\}}  \\ 
 \end{array}
  \label{eq:dOne}
\end{equation}

In order to expresses the joint survival indicator in terms of the survival of the bank and the counterparty, the third line of equation (\ref{eq:dOne}) is replaced in (\ref{eq:One_tau_sds}) and the result follows assuming that the joint default condition ($d{\bf 1}_{\left\{ {\tau _C  \le s} \right\}} d{\bf 1}_{\left\{ {\tau _B  \le s} \right\}}$) is neglected.

\begin{equation}
\begin{array}{l}
 {\bf 1}_{\left\{ {\tau  > s + ds} \right\}}  = {\bf 1}_{\left\{ {\tau _C  > s + ds} \right\}} {\bf 1}_{\left\{ {\tau _B  > s + ds} \right\}} \\
\;\;\;\;  = \left( {{\bf 1}_{\left\{ {\tau _C  > s} \right\}}  - d{\bf 1}_{\left\{ {\tau _C  \le s} \right\}} } \right)\left( {{\bf 1}_{\left\{ {\tau _B  > s} \right\}}  - d{\bf 1}_{\left\{ {\tau _B  \le s} \right\}} } \right)  \\ 
\;\;\;\;  \approx \left( {{\bf 1}_{\left\{ {\tau  > s} \right\}}  - {\bf 1}_{\left\{ {\tau_C  > s} \right\}} d{\bf 1}_{\left\{ {\tau _B  \le s} \right\}}  - {\bf 1}_{\left\{ {\tau_B  > s} \right\}} d{\bf 1}_{\left\{ {\tau _C  \le s} \right\}} } \right)  \\ 
 \end{array}
  \label{eq:One_tau_sds}
\end{equation}


\begin{equation}
\begin{array}{l}
 \xi _{s + ds}^S {\bf 1}_{\left\{ {\tau  > s} \right\}}  + (\xi _{s + ds}^C  - \xi _{s + ds}^S ){\bf 1}_{\left\{ {\tau_B > s} \right\}} d{\bf 1}_{\left\{ {\tau _C  \le s} \right\}}  \\ 
\;\;\;\;\;\;\;\;\;\;\;\;\;\;\;\;\;\;  + (\xi _{s + ds}^B  - \xi _{s + ds}^S ){\bf 1}_{\left\{ {\tau_C > s} \right\}} d{\bf 1}_{\left\{ {\tau _B  \le s} \right\}}  = 0 \\ 
 \end{array}
  \label{eq:FCsds3}
\end{equation}

Replacing equation (\ref{eq:One_tau_sds}) in (\ref{eq:FCsds}) yields equation (\ref{eq:FCsds3}) where the indicator functions no longer depend on $s+ds$. Now the dependence with respect to $s+ds$ is removed from $\xi_{s+ds}$ terms by using differentials. Replacing $\hat F_s = \hat V_s - M_s$ in $ \xi _{s + ds}^S$, see equation (\ref{eq:xiSCB}), adding and subtracting the terms $r_s^* M_s ds$ and $r_s^* \hat V_s ds$, where $r^*$ is the arbitrary rate of the principle of invariance and knowing that $d \hat V_s = \hat V_{s+ds} - \hat V_s$, yields equation (\ref{eq:xiS_sds}).

\begin{equation}
\xi _{s + ds}^S  = d\hat V_s  - r_s^* \hat V_s ds + dC_s  + M_s (r_s^*  - r_s^C )ds + \hat F_s (r_s^*  - r_s^F )ds
  \label{eq:xiS_sds}
\end{equation}

\begin{equation}
\begin{array}{l}
 (\xi _{s + ds}^C  - \xi _{s + ds}^S ){\bf 1}_{\left\{ {\tau_B  > s} \right\}} d{\bf 1}_{\left\{ {\tau _C  \le s} \right\}}  = (H_s^C  + M_s  - \hat V_s ){\bf 1}_{\left\{ {\tau_B  > s} \right\}} d{\bf 1}_{\left\{ {\tau _C  \le s} \right\}}  \\ 
 (\xi _{s + ds}^B  - \xi _{s + ds}^S ){\bf 1}_{\left\{ {\tau_C  > s} \right\}} d{\bf 1}_{\left\{ {\tau _B  \le s} \right\}}  = (H_s^B  + M_s  - \hat V_s ){\bf 1}_{\left\{ {\tau_C  > s} \right\}} d{\bf 1}_{\left\{ {\tau _B  \le s} \right\}}  \\ 
 \end{array}
  \label{eq:xiC_xiS_terms}
\end{equation}

\begin{equation}
\begin{array}{l}
 H_{s + ds}^C d{\bf 1}_{\left\{ {\tau _C  \le s} \right\}}  = (H_s^C  + dH_s^C )d{\bf 1}_{\left\{ {\tau _C  \le s} \right\}}  = H_s^C d{\bf 1}_{\left\{ {\tau _C  \le s} \right\}}  \\ 
 \hat V_{s + ds} d{\bf 1}_{\left\{ {\tau _C  \le s} \right\}}  = (\hat V_s  + d\tilde V_s )d{\bf 1}_{\left\{ {\tau _C  \le s} \right\}}  = \hat V_s d{\bf 1}_{\left\{ {\tau _C  \le s} \right\}}  \\ 
 ds \cdot d{\bf 1}_{\left\{ {\tau _C  \le s} \right\}}  = 0 \\ 
 dC_s \cdot d{\bf 1}_{\left\{ {\tau _C  \le s} \right\}}  = 0 \\ 
 \end{array}
  \label{eq:Inf_high_order}
\end{equation}

The rest of the terms of equation (\ref{eq:FCsds3}) are presented in equation (\ref{eq:xiC_xiS_terms}). They were obtained from equation (\ref{eq:xiSCB}) assuming the simplifications shown in equations (\ref{eq:Inf_high_order}). These simplifications are based on the fact that the product of two differential terms are negligible higher order infinitesimals. If these terms are replaced in equation (\ref{eq:FCsds3}), the final transformed expression of the funding condition is given by equation (\ref{eq:FCs3}).

\begin{equation}
\begin{array}{l}
 \xi _{s + ds}^S {\bf 1}_{\left\{ {\tau  > s} \right\}}  + (H_s^C  + M_s  - \hat V_s ){\bf 1}_{\left\{ {\tau_B > s} \right\}} d{\bf 1}_{\left\{ {\tau _C  \le s} \right\}}  \\ 
\;\;\;\;\;\;\;\;\;\;\;\;\;\;\;\;\;\;  + (H_s^B  + M_s  - \hat V_s ){\bf 1}_{\left\{ {\tau_C  > s} \right\}} d{\bf 1}_{\left\{ {\tau _B  \le s} \right\}}  = 0 \\ 
 \end{array}
  \label{eq:FCs3}
\end{equation}

\begin{equation}
\begin{array}{l}
 \int_t^\infty  {p_{t,s}^{r*} \left( {d\hat V_s  - r_s^* \hat V_s ds + dC_s } \right) {\bf 1}_{\left\{ {\tau  > s} \right\}} - p_{t,s}^{r*} \hat V_s d{\bf 1}_{\left\{ {\tau  \le s} \right\}}  + p_{t,s}^{r*} \hat V_s d{\bf 1}_{\left\{ {\tau  \le s} \right\}} } \\ 
  + \int_t^\infty  {p_{t,s}^{r*} M_s (r_s^*  - r_s^C ){\bf 1}_{\left\{ {\tau  > s} \right\}} ds + p_{t,s}^{r*} \hat F_s (r_s^*  - r_s^F ){\bf 1}_{\left\{ {\tau  > s} \right\}} ds}  \\ 
  + \int_t^\infty  {p_{t,s}^{r*} \left( {H_s^C  + M_s  - \hat V_s } \right){\bf 1}_{\left\{ {\tau_B > s} \right\}} d{\bf 1}_{\left\{ {\tau _C  \le s} \right\}} }  \\ 
  + \int_t^\infty  {p_{t,s}^{r*} \left( {H_s^B  + M_s  - \hat V_s } \right){\bf 1}_{\left\{ {\tau_C > s} \right\}} d{\bf 1}_{\left\{ {\tau _B  \le s} \right\}} }  = 0 \\ 
 \end{array}
  \label{eq:FCint_SCB}
\end{equation}

If equation (\ref{eq:xiS_sds}) is replaced in (\ref{eq:FCs3}), the resulting expression is multiplied by $p_{t,s}^{r*}$, the term $ p_{t,s}^{r*} \hat V_s d{\bf 1}_{\left\{ {\tau  \le s} \right\}}$ is added and subtracted and the whole expression is integrated from time $t$ to infinity, expression (\ref{eq:FCint_SCB}) is obtained.

\begin{equation}
\begin{array}{l}
 d{\bf 1}_{\left\{ {\tau  \le s} \right\}}  = {\bf 1}_{\left\{ {s < \tau  \le s + ds, \tau = \tau_C} \right\}} + {\bf 1}_{\left\{ {s < \tau \le s + ds, \tau = \tau_B} \right\}} + {\bf 1}_{\left\{ {s < \tau \le s + ds, \tau = \tau_C, \tau = \tau_B } \right\}} \\ 
\;\;\;\;\;\;\;\;\;\;\;\; \approx  {\bf 1}_{\left\{ {\tau_B > s} \right\}} d{\bf 1}_{\left\{ {\tau_C \le s } \right\}} + {\bf 1}_{\left\{ {\tau_C > s} \right\}} d{\bf 1}_{\left\{ {\tau_B \le s } \right\}}  \\ 
 \end{array}
  \label{eq:dOne_t_le_s}
\end{equation}

Taking expectation, ${\bf E}_t(\cdot)$, conditioned on time $t$ in equation (\ref{eq:FCint_SCB}) and replacing equation (\ref{eq:E_dexpV}) with $r=r^*$ on the first line of (\ref{eq:FCint_SCB}) would collapse the first, second and fourth terms of the integrand into $-\hat V_s {\bf 1}_{\left\{ {\tau  > s} \right\}}$. In addition, if the differential $d{\bf 1}_{\left\{ {\tau  \le s} \right\}}$ of the fifth term of the integrand is replaced by equation (\ref{eq:dOne_t_le_s})\footnote{The first to default indicator condition between $s$ and $s+ds$ is expressed as the sum of the first to default of the counterparty plus the first to default of the bank plus the joint default which is neglected. See equation (\ref{eq:OneFCsds}) to understand the final result.}, this term cancels the $-\hat V_s$ component inside the parenthesis of the third and fourth lines of equation (\ref{eq:FCint_SCB}), the CVA and DVA terms. Finally, the second line of equation (\ref{eq:FCint_SCB}) can be replaced by the collateral and funding adjustments of equation (\ref{eq:FColVA}). The final result is presented in equation (\ref{eq:FC_SCB3}).

\begin{equation}
\begin{array}{l}
  - \hat V_s {\bf 1}_{\left\{ {\tau  > s} \right\}} + {\bf E}_t \left[ {\int_t^\infty  {p_{t,s}^{r*} dC_s {\bf 1}_{\left\{ {\tau  > s} \right\}} } } \right] + ColVA_t^{r*}  + FVA_t^{r*}  \\ 
  + {\bf E}_t \left[ {\int_t^\infty  {p_{t,s}^{r*} \left( {H_s^C  + M_s } \right){\bf 1}_{\left\{ {\tau_B  > s} \right\}} d{\bf 1}_{\left\{ {\tau _C  \le s} \right\}} } } \right] \\ 
  + {\bf E}_t \left[ {\int_t^\infty  {p_{t,s}^{r*} \left( {H_s^B  + M_s } \right){\bf 1}_{\left\{ {\tau_C  > s} \right\}} d{\bf 1}_{\left\{ {\tau _B  \le s} \right\}} } } \right] = 0 \\ 
 \end{array}
  \label{eq:FC_SCB3}
\end{equation}

In order to calculate the second term of equation (\ref{eq:FC_SCB3}), $dC_s$ is expressed in terms of $V_s$ discounted by the rate $r_s^*$ through equation (\ref{eq:price_Cs}) applied to the rate $r^*$ and equation (\ref{eq:E_dexpV}) assuming $\tau \to \infty$. This yields equation (\ref{eq:dCs_VrF}):

\begin{equation}
dV_s^{r^* }  - r_s^* V_s^{r^* } ds + dC_s  = 0
  \label{eq:dCs_VrF}
\end{equation}

The calculation of the second term of equation (\ref{eq:FC_SCB3}) is shown in equation (\ref{eq:Et_dCs}) and it is obtained by replacing equation (\ref{eq:dCs_VrF}) in this second term and adding and subtracting $ p_{t,s}^{r*} V_s^{r*} d{\bf 1}_{\left\{ {\tau \le s} \right\}}$.

\begin{equation}
\begin{array}{l}
 {\bf E}_t \left[ {\int_t^\infty  {p_{t,s}^{r*} dC_s {\bf 1}_{\left\{ {\tau  > s} \right\}} } } \right] =  \\ 
 = - {\bf E}_t \left[ {\int_t^\infty  {p_{t,s}^{r*} (dV_s^{r*}  - r_s^* V_s^{r*} ds ){\bf 1}_{\left\{ {\tau  > s} \right\}} }  - p_{t,s}^{r*} V_s^{r*} d{\bf 1}_{\left\{ {\tau  \le s} \right\}}  + p_{t,s}^{r*} V_s^{r*} d{\bf 1}_{\left\{ {\tau  \le s} \right\}} } \right] \\ 
  =  V_t^{r*} {\bf 1}_{\left\{ {\tau  > t} \right\}}  - {\bf E}_t \left[ {\int_t^\infty  {p_{t,s}^{r*} V_s^{r*} d{\bf 1}_{\left\{ {\tau  \le s} \right\}} } } \right] \\ 
 \end{array}
  \label{eq:Et_dCs}
\end{equation}

Once more, replacing equation (\ref{eq:E_dexpV}) with $r=r^*$ in the second line of equation (\ref{eq:Et_dCs}) would collapse the first, second and third terms of the integrand into $- V_t^{r*} {\bf 1}_{\left\{ {\tau  > t} \right\}}$ yielding the third line of equation (\ref{eq:Et_dCs}).

\begin{equation}
\begin{array}{l}
 \hat V_t {\bf 1}_{\left\{ {\tau  > t} \right\}}  = V_t^{r*} {\bf 1}_{\left\{ {\tau  > t} \right\}}  + ColVA_t^{r*}  + FVA_t^{r*}  \\ 
\;\;\;\;\;\;\;\;\;\;\;\;\;\;\;\;\;  + {\bf E}_t \left[ {\int_t^\infty  {p_{t,s}^{r*} \left( {H_s^C  + M_s  - V_s^{r*} } \right){\bf 1}_{\left\{ {\tau_B > s} \right\}} d{\bf 1}_{\left\{ {\tau _C  \le s} \right\}} } } \right] \\ 
\;\;\;\;\;\;\;\;\;\;\;\;\;\;\;\;\;  + {\bf E}_t \left[ {\int_t^\infty  {p_{t,s}^{r*} \left( {H_s^B  + M_s  - V_s^{r*} } \right){\bf 1}_{\left\{ {\tau_C > s} \right\}} d{\bf 1}_{\left\{ {\tau _B  \le s} \right\}} } } \right] \\ 
 \end{array}
  \label{eq:VhatOne_tau_s}
\end{equation}

Finally, if equation (\ref{eq:dOne_t_le_s}) is replaced in (\ref{eq:Et_dCs}) and the result replaced in equation (\ref{eq:FC_SCB3}), the final expression (\ref{eq:VhatOne_tau_s}) equal to equation (\ref{eq:fundcr_eq}) to be proved follows.

\section{Improved solution for XVA calculation}
\label{sec:XVAimproved}

This appendix shows the hypotheses and transformations carried out to obtain equations (\ref{eq:fundcr_Vmx}) to (\ref{eq:CDVA_Vmx}) of section \ref{sec:NotCircularDepDefault} after solving the circular dependence and choosing $V_s^{r^{sys}}$ as exit price, from the general equations (\ref{eq:fundcr_eq}) to (\ref{eq:H_XVA}) of the invariance principle with default risk (see section \ref{sec:InvariancePrincipleDefault}).

The starting point is the equation (\ref{eq:fundcr_eq}) of the invariance principle with default risk. The rate $r^*$, which can be arbitrarily chosen without affecting the pricing equation (\ref{eq:fundcr_eq}), is set to the funding rate, $r^F$, so that the term $FVA_t^{r*}$ turns to zero. This breaks the circular dependence as the right hand side of equation (\ref{eq:Vhat_rF}) no longer depends on $\hat V$ and equation (\ref{eq:fundcr_eq}) becomes (\ref{eq:Vhat_rF}).

\begin{equation}
 \hat V_t {\bf 1}_{\left\{ {\tau  > t} \right\}}  = V_t^{r^F} {\bf 1}_{\left\{ {\tau  > t} \right\}}  + ColVA_t^{r^F}  + CVA_t^{r^F}  + DVA_t^{r^F}  \\ 
  \label{eq:Vhat_rF}
\end{equation}

If equation (\ref{eq:VrF_Vrsys_def}) is substituted in equation (\ref{eq:Vhat_rF}), the $ColVA_t^{r^F}$ term combines with the last adjustment term of equation (\ref{eq:VrF_Vrsys_def}) according to equation (\ref{eq:ColVA_Adj_rF_rsys}). See that the third line of this equation is obtained by adding and subtracting the integrand term ${p_{t,s}^{r^F } M_s } (r_s^{sys}  - r_s^F ){\bf 1}_{\left\{ {\tau  > s} \right\}} ds$ to the second line and taking into account that $F_s^{r^{sys}} = V_s^{r^{sys}} - M_s$. The fourth line of equation (\ref{eq:ColVA_Adj_rF_rsys}) is obtained by replacing the definitions given by equation (\ref{eq:FColVA_Vmx}). Replacing this calculation in equation (\ref{eq:Vhat_rF}) yields equation (\ref{eq:Vhat_rsys}).

\begin{equation}
\begin{array}{l}
 ColVA_t^{r^F }  + {\bf E}_t \left[ {\int_t^\infty  {p_{t,s}^{r^F } V_s^{r^{sys} } (r_s^{sys}  - r_s^F ){\bf 1}_{\left\{ {\tau  > s} \right\}} ds} } \right] =  \\ 
  = {\bf E}_t \left[ {\int_t^\infty  {p_{t,s}^{r^F } M_s (r_s^F  - r_s^C ){\bf 1}_{\left\{ {\tau  > s} \right\}} ds + p_{t,s}^{r^F } V_s^{r^{sys} } (r_s^{sys}  - r_s^F ){\bf 1}_{\left\{ {\tau  > s} \right\}} ds} } \right] \\ 
  = {\bf E}_t \left[ {\int_t^\infty  {p_{t,s}^{r^F } M_s (r_s^{sys}  - r_s^C ){\bf 1}_{\left\{ {\tau  > s} \right\}} ds + p_{t,s}^{r^F } F_s^{r^{sys} } (r_s^{sys}  - r_s^F ){\bf 1}_{\left\{ {\tau  > s} \right\}} ds} } \right] \\ 
  = - ColVA_t^{sys}  - FVA_t^{sys}  \\ 
 \end{array}
  \label{eq:ColVA_Adj_rF_rsys}
\end{equation}

\begin{equation}
 \hat V_t {\bf 1}_{\left\{ {\tau  > t} \right\}}  = V_t^{r^{sys}} {\bf 1}_{\left\{ {\tau  > t} \right\}}  - ColVA_t^{sys} - FVA_t^{sys} + CVA_t^{r^F}  + DVA_t^{r^F}  \\ 
  \label{eq:Vhat_rsys}
\end{equation}

\begin{equation}
\begin{array}{l}
 CVA_t^{r^F}  = {\bf E}_t \left[ {\int_t^\infty  {p_{t,s}^{r^F } \left( {H_s^C  + M_s  - V_s^{r^F} } \right){\bf 1}_{\left\{ {\tau  > s} \right\}} d{\bf 1}_{\tau _C  \le s} } } \right] \\ 
\;\;\;\;  = {\bf E}_t \left[ {\int_t^\infty  {p_{t,s}^{r^F } \left( {H_s^C  + M_s  - V_s^{r^{sys}} } \right){\bf 1}_{\left\{ {\tau  > s} \right\}} d{\bf 1}_{\tau _C  \le s} } } \right] = \\  
\;\;\;\;  =  - {\bf E}_t \left[ {\int_t^\infty  {p_{t,s}^{r^F } (1 - R_s^C )(V_s^{r^{sys} }  - M_s )^ +  {\bf 1}_{\left\{ {\tau  > s} \right\}} d{\bf 1}_{\tau _C  \le s} } } \right] =  - CVA_t^{sys}  \\ 
 \end{array}
  \label{eq:CVArF}
\end{equation}

\begin{equation}
\begin{array}{l}
 DVA_t^{r^F}  = {\bf E}_t \left[ {\int_t^\infty  {p_{t,s}^{r^F } \left( {H_s^B  + M_s  - V_s^{r^F} } \right){\bf 1}_{\left\{ {\tau  > s} \right\}} d{\bf 1}_{\tau _B  \le s} } } \right] \\ 
 \;\;\;\;  = {\bf E}_t \left[ {\int_t^\infty  {p_{t,s}^{r^F } \left( {H_s^B  + M_s  - V_s^{r^{sys}} } \right){\bf 1}_{\left\{ {\tau  > s} \right\}} d{\bf 1}_{\tau _B  \le s} } } \right] \\ 
\;\;\;\;  =  - {\bf E}_t \left[ {\int_t^\infty  {p_{t,s}^{r^F } (1 - R_s^B )(V_s^{r^{sys} }  - M_s )^ -  {\bf 1}_{\left\{ {\tau  > s} \right\}} d{\bf 1}_{\tau _B  \le s} } } \right] =  - DVA_t^{sys}  \\ 
 \end{array}
  \label{eq:DVArF}
\end{equation}

Setting the rate $r^* = r^F$ and the exit price $\tilde V_s  = V_s^{r^{sys} }$ for the CVA and DVA terms of equation (\ref{eq:Vhat_rsys}) and replacing $V_s^{r^F}$ in terms of $V_s^{r^{sys}}$ according to equation (\ref{eq:VrF_Vrsys_def}) yields equations (\ref{eq:CVArF}) and (\ref{eq:DVArF}). See that when $V_s^{r^F}$ is replaced according to equation (\ref{eq:VrF_Vrsys_def}), the term $V_t^{r^{sys}}$ continues to appear within the parenthesis of the second line of equations (\ref{eq:CVArF}) and (\ref{eq:DVArF}) but the adjustment term disappears. Equation (\ref{eq:Adj_du_dOne}) shows that the contribution of this adjustment term for the CVA equation is zero due to the fact that $d{\bf 1}_{\left\{ {\tau _C  \le s} \right\}}$ is measurable with respect to time $s$ getting inside the expectation and the product of differentials, $dsd{\bf 1}_{\tau _C  \le s}$, is negligible. The transformation of the terms inside the parenthesis of the second line of equation (\ref{eq:CVArF}) are explained by equation (\ref{eq:HC_M_Vsys}), where the term $(V_s^{r^{sys}}  - M_s )^+$ is added and subtracted so that combining positive and negative parts cancels the term $M_s  - V_s^{r^{sys}}$. The same approach is carried out for the DVA equation. The final expression for the parenthesis inside the CVA and DVA integrands is presented in equation (\ref{eq:HCB_M_Vsys}).

\begin{equation}
{\bf E}_t \left( {\int_t^\infty  {{\bf E}_s \left[ {\int_s^\infty  {p_{s,u}^{r^F } V_u^{sys} (r_u^{sys}  - r_u^F ){\bf 1}_{\left\{ {\tau  > u} \right\}} du} } \right]}  \cdot d{\bf 1}_{\left\{ {\tau _C  \le s} \right\}} } \right) = 0
  \label{eq:Adj_du_dOne}
\end{equation}

\begin{equation}
\begin{array}{l}
 H_s^C  + M_s  - V_s^{r^{sys} }  = R_s^C (V_s^{r^{sys} }  - M_s )^ +   + (V_s^{r^{sys} }  - M_s )^ -   \\ 
\;\;\;\;\;\;\;\;\;  + (V_s^{r^{sys} }  - M_s )^ +   - (V_s^{r^{sys} }  - M_s )^ +   + M_s  - V_s^{r^{sys} }  \\ 
 \end{array}
  \label{eq:HC_M_Vsys}
\end{equation}

\begin{equation}
\begin{array}{l}
 H_s^C  + M_s  - V_s^{r^{sys} }  =  - (1 - R_s^C )(V_s^{r^{sys} }  - M_s )^ +   \\ 
 H_s^B  + M_s  - V_s^{r^{sys} }  =  - (1 - R_s^C )(V_s^{r^{sys} }  - M_s )^ -   \\ 
 \end{array}
  \label{eq:HCB_M_Vsys}
\end{equation}

\begin{equation}
 \hat V_t {\bf 1}_{\left\{ {\tau  > t} \right\}}  = V_t^{r^{sys}} {\bf 1}_{\left\{ {\tau  > t} \right\}}  - ColVA_t^{sys} - FVA_t^{sys} - CVA_t^{sys}  - DVA_t^{sys}  \\ 
  \label{eq:Vhat_sys}
\end{equation}

Replacing equations (\ref{eq:CVArF}) and (\ref{eq:DVArF}) in equation (\ref{eq:Vhat_rsys}) yields the expression (\ref{eq:Vhat_sys}) to be proved, equal to equation (\ref{eq:fundcr_Vmx}).

%
%

%
%
%
%
%
%
%
%
%
%
%
%

\end{document}